\documentclass[12pt,twoside]{article}
\usepackage[mathscr]{eucal}
\usepackage{amsmath,amsfonts,amssymb,amsthm}
\usepackage[matrix,arrow]{xy}
\usepackage{times}
\usepackage{pdfsync}
\usepackage{graphicx}
\usepackage{epstopdf}
\usepackage{multibox}
\usepackage{dcolumn}
\usepackage{hyperref}

\def\d_Vphi{\text{d}_V\hspace{-0.06em}\phi}
\def\d_Vphibar{\text{d}_V\hspace{-0.06em}\bar\phi}
\def\d_Vxi{\text{d}_V\hspace{-0.06em}\xi}

\def\be{\begin{eqnarray}}
\def\ee{\end{eqnarray}}
\def\beann{\begin{eqnarray*}}
\def\eeann{\end{eqnarray*}}
\def\beq{\begin{equation}}
\def\eeq{\end{equation}}
\def\ba{\begin{array}}
\def\ea{\end{array}}
\def\ben{\begin{enumerate}}
\def\een{\end{enumerate}}
\def\bea{\begin{eqnarray}}
\def\eea{\end{eqnarray}}

\def\5{\bar }
\def\6{\partial }
\def\7{\hat }
\def\4{\tilde }
\advance\textheight\topskip
\renewcommand{\tilde}{\widetilde}
\renewcommand{\hat}{\widehat}

\newcommand{\bref}[1]{\textbf{\ref{#1}}}

\newcommand{\dd}{\partial}
\renewcommand{\d}{\partial}

\newcommand{\binner}[2]{%
  {\langle}\kern-4.15pt{\langle}#1{,}\,#2{\rangle}\kern-4.15pt{\rangle}}

\newcommand{\half}{\frac{1}{2}}

\newcommand{\ffrac}[2]{\raisebox{.5pt}%
  {\footnotesize$\displaystyle\frac{#1}{#2}$}\kern1pt}

\newcommand{\dover}[2]{\ffrac{\dd #1}{\dd #2}}

\def\cA{\mathcal{A}}

\def\cK{\mathcal{K}}
\def\cL{\mathcal{L}}

\def\cO{\mathcal{O}}

\def\cQ{\mathcal{Q}}
\def\cR{\mathcal{R}}

\numberwithin{equation}{section} \makeatletter
\@addtoreset{equation}{section}

\begin{document}

\def\mytitle{Aspects of the BMS/CFT correspondence}

\pagestyle{myheadings} \markboth{\textsc{\small Barnich, Troessaert}}{%
  \textsc{\small BMS/CFT correspondence}} \addtolength{\headsep}{4pt}

\begin{flushright}\small
ULB-TH/09-28\end{flushright}

\begin{centering}

  \vspace{1cm}

  \textbf{\Large{\mytitle}}



  \vspace{1.5cm}

  {\large Glenn Barnich$^{a}$ and C\'edric Troessaert$^{b}$}

\vspace{.5cm}

\begin{minipage}{.9\textwidth}\small \it \begin{center}
   Physique Th\'eorique et Math\'ematique\\ Universit\'e Libre de
   Bruxelles\\ and \\ International Solvay Institutes \\ Campus
   Plaine C.P. 231, B-1050 Bruxelles, Belgium \end{center}
\end{minipage}

\end{centering}

\vspace{1cm}

\begin{center}
  \begin{minipage}{.9\textwidth}
    \textsc{Abstract}. After a review of symmetries and classical
    solutions involved in the ${\rm AdS_3/CFT_2}$ correspondence, we
    apply a similar analysis to asymptotically flat spacetimes at null
    infinity in $3$ and $4$ dimensions. In the spirit of two
    dimensional conformal field theory, the symmetry algebra of
    asymptotically flat spacetimes at null infinity in 4 dimensions is
    taken to be the semi-direct sum of supertranslations with
    infinitesimal local conformal transformations and not, as usually
    done, with the Lorentz algebra. As a first application, we derive
    how the symmetry algebra is realized on solution space. In
    particular, we work out the behavior of Bondi's news tensor, mass
    and angular momentum aspects under local conformal
    transformations.
  \end{minipage}
\end{center}

\vfill

\noindent
\mbox{}
\raisebox{-3\baselineskip}{%
  \parbox{\textwidth}{\mbox{}\hrulefill\\[-4pt]}}
{\scriptsize$^a$Research Director of the Fund for Scientific
  Research-FNRS. E-mail: gbarnich@ulb.ac.be\\$^b$ Research Fellow of
  the Fund for Scientific Research-FNRS. E-mail: ctroessa@ulb.ac.be}

\thispagestyle{empty}
\newpage

\begin{small}
{\addtolength{\parskip}{-1.5pt}
 \tableofcontents}
\end{small}
\newpage

\section{Introduction}
\label{sec:introduction}

When studying dualities between string or gravitational theories in
anti-de Sitter backgrounds and conformal theories on their boundaries
\cite{Maldacena:1997re}, the $3$-dimensional case is especially
interesting. Indeed, the symmetries of asymptotically $AdS$
spacetimes \cite{Henneaux:1985tv,Brown:1986nw,Henneaux:1985ey} provide
a representation of the algebra of conformal Killing vectors of the
flat boundary metric. In $3$ spacetime dimensions, this algebra is
infinite-dimensional and powerful techniques of $2$-dimensional
conformal field theory are available.

Historically, the first example where the asymptotic symmetry group is
enhanced with respect to the isometry group of the background metric
and becomes infinite-dimensional is the one of asymptotically flat
$4$-dimensional spacetimes at null infinity
\cite{Bondi:1962px,Sachs:1962wk,Sachs2}.  In this case, the induced
metric is $2$-dimensional because the boundary is a null surface. The
asymptotic symmetry group of non singular transformations is the
well-known Bondi-Metzner-Sachs group. It consists of the semi-direct
product of the group of globally defined conformal transformations of
the unit $2$-sphere, which is isomorphic to the orthochronous
homogeneous Lorentz group, times the infinite-dimensional abelian
normal subgroup of so-called supertranslations.

There is a further enhancement when one focuses on infinitesimal
transformations and does not require the associated finite
transformations to be globally well-defined. The symmetry algebra is
then the semi-direct sum of the infinitesimal local conformal
transformations of the $2$-sphere with the abelian ideal of
supertranslations, and now both factors are infinite-dimensional
\cite{Barnich:2009se}.

A first hint on what the symmetry algebra of asymptotically flat
spacetimes at null infinity might be, independently of precise
fall-off conditions on the metric, can be obtained by solving the
Killing equations to leading order. This has been done in
\cite{Barnich:2006avcorr} in all dimensions greater than $3$. In $4$
dimensions, the infinite-dimensional nature of the conformal factor of
the $\mathfrak{bms}_4$ algebra has been emphasized. In $3$ dimensions,
the asymptotic symmetry algebra $\mathfrak{bms}_3$, originally derived
in \cite{Ashtekar:1996cm,Ashtekar:1996cd}, has been recovered. The
algebra of surface charges has been shown to provide a centrally
extended representation of $\mathfrak{bms}_3$ which has been related
by a contraction, similiar to that from $\mathfrak{so}(2,2)$ to
$\mathfrak{iso}(2,1)$, to the centrally extended Poisson bracket
algebra of surface charges of asymptotically anti-de Sitter spacetimes
in $3$ dimensions.

The aim of the present work is to reconsider from the point of view of
local conformal transformations the $4$-dimensional case which is, in
some sense at least, of direct physical relevance. In particular, we
provide a detailed derivation of the natural generalization of the
$\mathfrak{bms}_4$ algebra discussed above. No modification of well
studied boundary conditions is needed and the transformations are
carefully distinguished from conformal rescalings. 

A major motivation for our investigation comes from Strominger's
derivation \cite{Strominger:1998eq} of the Bekenstein-Hawking entropy
for black holes that have a near horizon geometry that is locally
$AdS_3$ by using the Brown-Henneaux analysis of the surface charge
algebra of asymptotically anti-de Sitter spacetimes at spatial
infinity. More recently, a similar analysis has been used to derive
the Bekenstein-Hawking entropy of an extreme 4-dimensional Kerr black
hole \cite{Guica:2008mu}. One of our hopes is to make progress along
these lines in the non extreme case, either directly from an analysis
at null infinity or by making a similar analysis at the horizon, as
discussed previously for instance in
\cite{Carlip:1994gy,Solodukhin:1998tc,Carlip:1998wz,Carlip:1999cy,%
  Park:1999tj,Park:2001zn,Sachs:2001qb,Koga:2001vq,Carlip:2002be,%
Silva:2002jq,Kang:2004js,Koga:2006ez,Koga:2006sb}.

Related work includes \cite{Ashtekar:1981sf,Ashtekar:1987tt} on
asymptotic quantization where for instance the implications of
supertranslations for the gravitational $S$-matrix have been
discussed. Asymptotically flat spacetimes at null infinity in higher
spacetime dimensions have been investigated for instance in
\cite{Hollands:2003ie,Hollands:2003xp,Hollands:2004ac,Tanabe:2009va},
while various aspects of holography in $4$ dimensions have been
studied in some details
in~\cite{Susskind:1998vk,Polchinski:1999ry,deBoer:2003vf,%
  Arcioni:2003xx,Arcioni:2003td,Solodukhin:2004gs,Gary:2009mi}. In
particular, a symmetry algebra of the kind that we derive and study
here has been conjectured in \cite{Banks:2003vp}.

This paper is organized as follows. Asymptotically $AdS_3$ spacetimes
in Fefferman-Graham form are briefly reviewed in the next section in
order to illustrate the approach adopted for the asymptotically flat
case in a well-studied situation. In order to be self-contained, we
include details of the computations. The section is based on results
originally derived and discussed for instance in
\cite{fefferman:1985,Brown:1986nw,graham:1991,Henningson:1998gx,%
  1999AIPC..484..147B,Skenderis:1999nb,Graham:1999jg,Imbimbo:1999bj,%
  Rooman:1999km,Bautier:2000mz,Papadimitriou:2005ii}: in particular,
for asymptotically $AdS_3$ spacetimes in the sense of
Fefferman-Graham, the gauge is completely fixed as all subleading
orders of the asymptotic Killing vectors are uniquely determined; 
furthermore, the general solution of the equations of motion can be
written down in closed form for an arbitary boundary metric. We then
compute how the local conformal algebra in 2 dimensions is realized on
solution space and analyze how this realization is probed by a
covariant version of the Brown-Henneaux charge algebra.

A novel result in this context concerns the representation of the
local conformal algebra and its abelian extension by conformal
rescalings in terms of spacetime vectors equipped with a new, modified
Dirac-type Lie bracket taking into account the metric dependence of
the asymptotic Killing vectors due to the gauge fixing.

In section~\bref{sec:bms3}, we re-analyze asymptotically flat
spacetimes at null infinity in $3$ dimensions. We start by rederiving
their symmetries in the form suggested by the analysis of the $4$
dimensional case by Sachs and by the Fefferman-Graham analysis of
asymptotically $AdS$ spacetimes in $3$ dimensions. In particular, we
discuss the extended symmetry algebra and its representation in terms
of spacetime vectors equipped with the modified Lie
bracket. Furthermore, we provide the general solution to the equations
of motions.  We then analyze how $\mathfrak{bms}_3$ is realized on
solution space and how this realization is probed by the centrally
extended covariant surface charge algebra.

In the main section~\bref{sec:asympt-flat-spac} on asymptotically flat
spacetimes at null infinity in $4$ spacetime dimensions, we derive the
asymptotic algebra and prove the main result on local conformal
transformations announced in \cite{Barnich:2009se}. In order to do so,
we follow closely the original analysis of Bondi, Metzner and Sachs
\cite{Bondi:1962px,Sachs:1962wk,Sachs2}, but as suggested by the
geometrical analysis of Penrose \cite{PhysRevLett.10.66}, we allow for
an arbitrary conformal factor in the boundary metric. More precisely,
by allowing meromorphic functions on the Riemann sphere, we show that
the asymptotic symmetry algebra involves instead of the Lorentz
algebra two copies of the Witt algebra that act in a natural way on
the abelian ideal of supertranslations.  By restricting ourselves to
exact isometries of Minkowski spacetime, we identify the Poincar\'e
subalgebra of $\mathfrak{bms}_4$. We show how the asymptotic symmetry
algebra is represented by the spacetime vectors equipped with the
modified Lie bracket and generalize these results to the extended
algebra including conformal rescalings of the boundary metric.

We then turn to a detailed discussion of solution space by mostly
following \cite{Tamburino:1966,Winicour:1985pi}. Under appropriate
assumptions, the arbitrary functions which arise as integration
constants in the general solution to the equations of motion are
identified. They involve in particular the so-called news tensor
together with the mass and angular momentum aspects.  We include a
discussion of the leading logarithmic term and generalize existing
results by allowing the conformal factor to explicitly depend on time.

The second main result consists in deriving the transformation
properties under the asymptotic symmetry algebra $\mathfrak{bms}_4$ of
some of the fields characterizing the solutions. In particular,
conformal dimensions and central charges under local conformal
transformations of the news tensor, the mass and angular momentum
aspects are worked out. 

\section{Asymptotically ${\rm \bf AdS_3}$ spacetimes in
  Fefferman-Graham form}
\label{sec:poinc-gener-pauli}

The Fefferman-Graham form for the line element of a $3$
dimensional asymptotically anti-de Sitter spacetime is
\begin{equation}
ds^2 = \frac{l^2}{r^2} dr^2 + g_{AB}(r,x^C)\, dx^Adx^B,\label{eq:GF}
\end{equation}
with $g_{AB} = r^2\bar\gamma_{AB}(x^C)+ O(1)$, where $\bar\gamma_{AB}$
is a conformally flat $2$-dimensional metric.  For explicit
computations we will sometimes choose the parametrization
$\bar\gamma_{AB}=e^{2\varphi} \eta_{AB}$ with $\varphi(x^C)$ and
$\eta_{AB}$ the flat metric on the cylinder, $\eta_{AB}
dx^Adx^B=-d\tau^2+d\phi^2$, $\tau=\frac{1}{l}t$.

\subsection{Asymptotic symmetries}
\label{sec:asympt-symma}

The transformations leaving this form of the metric invariant are
generated by vector fields satisfying
\begin{equation}
\cL_\xi g_{rr}=0=\cL_\xi g_{rA},\label{eq:ads31}\quad
\cL_\xi g_{AB}=O(1),
\end{equation}
which implies
\begin{eqnarray}
\left\{\begin{array}{l}\xi^r =-\half\psi r, \\ \xi^A  =  Y^A +
    I^A,\quad 
 I^A=-\frac{l^2}{2}\d_B
  \psi\int_r^\infty \frac{dr^\prime}{r^\prime}
  g^{AB}=-\frac{l^2}{4r^2}
\bar\gamma^{AB}\d_B
  \psi+O(r^{-4}),
\end{array}\right.\label{eq:FGvect}
\end{eqnarray}
where $Y^A$ is a conformal Killing vector of $\bar\gamma_{AB}$, and
thus of $\eta_{AB}$, while 
$\psi=\bar D_A Y^A$ is the conformal factor. 

Indeed, the inverse to metric \eqref{eq:GF} is
\begin{equation*}
g^{\mu\nu} = \left(\begin{array}{cc}
 \frac{r^2}{l^2} & 0 \\
0 & g^{AB}
\end{array}\right)
\end{equation*}
where $g^{AB}g_{BC}=\delta^A_C$. From $\cL_\xi g_{rr}=0$, we find
$\xi^r= A r$ for some $A(x^C)$. From $\cL_\xi g_{rA} = 0$ we find
$\partial_r \xi^A=-g^{AB}\frac{l^2}{r}\partial_B A$ so that $\xi^A =
Y^A + I^A$ for some $Y^A(x^C)$ and where $I^A = {l^2}\partial_B A
\,\int_r^\infty dr' \, g^{AB}r'^{-1}$. Finally, $\cL_\xi
g_{AB}=O(1)$ requires $Y^A$ to be a conformal Killing
vector of $\bar\gamma_{AB}$ and $A=-\half \psi$.

Let $\hat Y^A=[Y_1,Y_2]^A$, $\hat
\psi=\bar D_A \hat Y^A$, denote by
$\delta^g_{\xi_1}\xi^\mu_2$ the change induced in $\xi^\mu_2(g)$ due
to the variation $\delta^g_{\xi_1}g_{\mu\nu}=\cL_{\xi_1}g_{\mu\nu}$
and define
\begin{equation}
  \label{eq:44a}
 [\xi_1,\xi_2]^\mu_M =[\xi_1,\xi_2]^\mu-\delta^g_{\xi_1}\xi^\mu_2+
\delta^g_{\xi_2}\xi^\mu_1. 
\end{equation}
For vectors $\xi_1,\xi_2$ given in \eqref{eq:FGvect}, we have
\begin{equation*}
  [\xi_1,\xi_2]^r_M=-\half \hat\psi r,\quad [\xi_1,\xi_2]^A_M=\hat
  Y^A+\hat I^A,
\end{equation*}
where $\hat I^A$ denotes $I^A$ with $\psi$ replaced by $\hat
\psi$. 

Indeed, for the $r$ component, we have $\delta^g_{\xi_1}\xi^r_2=0$ and
the result follows by direct computation of the Lie
bracket. Similarily, $\lim_{r\to\infty}[\xi_1,\xi_2]^A_M=\hat Y^A$.
Finally, using $\d_r\xi^r=\frac{1}{r}\xi^r$ and
$\d_r\xi^A=-\frac{l^2}{r^2}\d_B\xi^r g^{BA}$ a straightforward
computation shows that
$\d_r([\xi_1,\xi_2]^A_M)=-\frac{l^2}{r^2}\d_B([\xi_1,\xi_2]^r_M) g^{BA}$, which gives
the result. It thus follows that on an asymptotically anti-de Sitter
spacetime in the sense of Fefferman-Graham (solving or not Einstein's
equations with cosmological constant):

{\em The spacetime vectors \eqref{eq:FGvect}
  equipped with the bracket $[\cdot,\cdot]_M$ form a faithful
  representation of the conformal algebra.}

By conformal algebra, we
mean here the direct sum of 2 copies of the Witt algebra.
Furthermore, since $\delta^g_{\xi_1}\xi^r_2=0$,
$\delta^g_{\xi_1}\xi^A_2=O(r^{-4})$, it follows that these vectors
form a representation of the conformal algebra only up to terms of
order $O(r^{-4})$ when equipped with the standard Lie bracket.

In terms of light-cone coordinates, 
$x^\pm=\tau\pm\phi$,
$2\d_\pm=\dover{}{\tau}\pm\dover{}{\phi}$, we have 
$\bar\gamma_{AB}dx^Adx^B=-e^{2\varphi}dx^+dx^-$, and if,
\begin{eqnarray}
Y^\pm(x^\pm)\d_\pm=\sum_{n\in \mathbf Z} c^n_{\pm} l^\pm_n,\quad
l^\pm_n=-(x^\pm)^{n+1}\d_{\pm} , 
\end{eqnarray}
the algebra in terms of the basis vectors $l^\pm_n$ reads 
\begin{equation}
  \label{eq:13}
[l^\pm_m,l^\pm_n]=(m-n)l^\pm_{m},\quad  [l^\pm_m,l^\mp_n]=0.
\end{equation}

More generally, one can also consider the transformations that leave
the Fefferman-Graham form of the metric invariant up to a Weyl
rescaling of the boundary metric $\bar\gamma_{AB}$. They are generated
by spacetime vectors such that
\begin{equation}
\cL_\xi g_{rr}=0=\cL_\xi g_{rA},\label{eq:ads33}\quad 
\cL_\xi g_{AB}=2\omega g_{AB}+O(1). 
\end{equation}
It is then straightforward to see that the general solution is given
by the vectors \eqref{eq:FGvect}, where $\psi$ is replaced by
$\tilde\psi=\psi-2\omega$. When equipped with the modified Lie bracket
$[\cdot,\cdot]_M$ these vectors now form a faithful representation of
the extension of the two dimensional conformal algebra defined by
elements $(Y,\omega)$ and the Lie bracket $(\hat
Y,\hat\omega)=[(Y_1,\omega_1),(Y_2,\omega_2)]$,
\begin{equation}
  \label{eq:45}
  \hat Y^A=Y^B_1\d_B Y_2^A-Y^B_2\d_B Y_1^A,\quad \hat\omega=0,
\end{equation}
with $\omega(x^C)$ arbitrary and $Y^A$ conformal Killing vectors of
$\bar\gamma_{AB}$ and thus also of $\eta_{AB}$. The asymptotic
symmetry algebra is then the direct sum of the abelian ideal of
elements of the form $(0,\omega)$ and of 2 copies of the Witt algebra.

Indeed, we have $\lim_{r\to\infty}(\frac{1}{r}[\xi_1,\xi_2]^r_M)=
-\half Y^A_1\d_A\tilde\psi_2+\d_C\omega_1 Y^C_2+(1\leftrightarrow
2)=-\half \hat \psi$ and $\d_r(\frac{1}{r}[\xi_1,\xi_2]^r_M)=0$, while
the proof for the $A$-component is unchanged. 

\subsection{Solution space}
\label{sec:solution-spacea}

Let us now start with an arbitrary metric of the form \eqref{eq:GF},
without any assumptions on the behavior in $r$ and let $k^A_B=\half
g^{AC}g_{CB,r}$. One can then define $K^A_B$ through the relation
$k^A_B=\frac{1}{r}\delta^A_B +\frac{1}{r^3} K^A_B$.
We have
\begin{equation*}
\begin{gathered}
\Gamma^r_{rr}   =   -\frac{1}{r},\quad 
\Gamma^r_{rA}  =  0,\quad \Gamma^A_{rr}  =  0,\\
\Gamma^r_{AB}  = -\frac{r^2}{l^2} k_{AB},\quad
\Gamma^A_{rB} =k^A_B,\quad 
\Gamma^A_{BC}  = {}^{(2)}\Gamma^A_{BC}, 
\end{gathered}
\end{equation*}
where $ {}^{(2)}\Gamma^A_{BC}$ denotes the Christoffel symbol
associated to the $2$-dimensional metric $g_{AB}$, which is used to
lower indices on $k^A_B$. If ${K^{TA}}_B$ denotes the traceless part of
$K^{A}_B$, the equations of motion are organized as follows
\begin{gather}
  \label{eq:80a}
  g^{AB}G_{AB}-\frac{2}{l^2} =0\Longleftrightarrow
  \d_r K=-r^{-3}(\half K^2+{K^{T}}^A_B{K^{T}}^B_A), \\\label{eq:80b}
G_{AB}-\half g_{AB}g^{CD}G_{CD}=0\Longleftrightarrow
\d_r {K^{T}}^A_B=-r^{-3} K {K^{T}}^A_B, \\\label{eq:80c}
G_{rA}\equiv r^{-3}( {}^{(2)}D_B K^B_A-\d_A K)=0,\\\label{eq:80d}
G_{rr}-\frac{1}{l^2}g_{rr}\equiv \half\big[r^{-6} (\half K^2
-{K^{T}}^A_B{K^{T}}^B_A )+2 r^{-4} K -\frac{l^2}{r^2}\, {}^{(2)}R\big]=0.
\end{gather}
Combining the Bianchi identities $2(\sqrt{-g}
G^\beta_\alpha)_{,\beta}+\sqrt{-g}
G_{\beta\gamma}{g^{\beta\gamma}}_{,\alpha}\equiv 0$ with the covariant
constancy of the metric, we get the identities 
\begin{multline}
  \label{eq:81}
  2(\frac{r}{l}\sqrt{|{}^{(2)}g|}G_{rA})_{,r}+2(\frac{l}{r}\sqrt{|{}^{(2)}g|}g^{BC}
  [G_{CA}-\frac{1}{l^2} g_{CA}])_{,B}
  +\\+\frac{l}{r}\sqrt{|{}^{(2)}g|}(G_{BC}-\frac{1}{l^2}
  g_{BC}){g^{BC}}_{,A}\equiv 0,
\end{multline}
\begin{multline}
  \label{eq:82}
\big(\frac{r}{l}\sqrt{|{}^{(2)}g|}
[G_{rr}-\frac{1}{r^2}]\big)_{,r} 
+(\frac{l}{r}\sqrt{|{}^{(2)}g|}g^{BA}
G_{Ar})_{,B}+\\+ \frac{1}{l}\sqrt{|{}^{(2)}g|}
[G_{rr}-\frac{1}{r^2}]-\frac{l}{r}\sqrt{|{}^{(2)}g|}(G_{AB}-\frac{1}{l^2}
g_{AB} )k^{AB}\equiv 0.
\end{multline}

To solve the equations of motion, we first contract \eqref{eq:80b}
with ${K^{T}}^B_A$, which gives
\begin{equation*}
    \d_r ({K^{T}}^A_B{K^{T}}^B_A)=-2r^{-3} K {K^{T}}^A_B{K^{T}}^B_A.
\end{equation*}
If we assume ${K^{T}}^A_B{K^{T}}^B_A=\half \cK^2$, we can take the sum and
difference with \eqref{eq:80a} to get 
\begin{equation*}
  \d_r(K+\cK)=-\frac{1}{2}r^{-3}(K+\cK)^2,\quad 
\d_r(K-\cK)=-\frac{1}{2}r^{-3}(K-\cK)^2, 
\end{equation*}
which can be solved in terms of $2$ integration ``constants'' $C(x^B),D(x^B)$
\begin{equation*}
  K=-\frac{1}{C+\half r^{-2}}-\frac{1}{D+\half r^{-2}},\quad 
{K^{T}}^A_B{K^{T}}^B_A=\frac{(D-C)^2}{2(C+\half r^{-2})^2(D+\half r^{-2})^2}.
\end{equation*}
When used in \eqref{eq:80b}, we find 
\begin{equation*}
  {K^{T}}^A_B={A^{T}}^A_B(\frac{1}{C+\half r^{-2}}-\frac{1}{D+\half r^{-2}}),\quad
  {A^{T}}^A_B{A^{T}}^B_A=\half,
\end{equation*}
and can now reconstruct the metric from the equation $\d_r g_{AB}=2
g_{AC} k^C_B$. Defining  $\Theta=\frac{1}{D}+\frac{1}{C}$,
$\Omega=\frac{1}{D}-\frac{1}{C}$, we get 
\begin{equation}
  \label{eq:88}
  g_{AB}=
r^2\bar\gamma_{AB}\big[1+\frac{1}{2r^2}
  \Theta
  +\frac{1}{16r^4}(\Theta^2+\Omega^2)\big]
  +A^T_{AB}\big[\Omega
  +\frac{1}{4r^2}\Theta\Omega\big],
\end{equation}
where $\bar\gamma_{AB}$ are additional integration constants,
restricted by the condition that $\bar\gamma_{AB}$ is symmetric, of
signature $-1$. The index on $A^{TA}_B$ is lowered with
$\bar\gamma_{AB}$, with $A^T_{AB}$ requested to be symmetric. It
follows that ${A^{T}}^A_B$ contains only $1$ additional independent
integration constant.  Writing
$g_{AB}=r^2\bar\gamma_{AB}+\gamma_{AB}$, with
$\gamma_{AB}=\hat\gamma_{AB}+ o(r^0)$, we have $K^A_B=-\hat
\gamma^A_B+o(r^0)$ where the index on $\hat\gamma^A_B$ has been lifted
with $\bar\gamma^{AB}$, the inverse of $\bar\gamma_{AB}$.

When \eqref{eq:80a} and \eqref{eq:80b} are satisfied, the Bianchi
identity \eqref{eq:81} implies that
$r\sqrt{|{}^{(2)}g|}G_{rA}$ does not depend on $r$. The equation of
motion \eqref{eq:80c} then reduces to the condition 
\begin{equation}
  \label{eq:87}
  \bar D_B \hat \gamma^B_A-\d_A \hat \gamma=0,
\end{equation}
where $\hat \gamma=\hat\gamma^A_A$.  When this condition holds in
addition to \eqref{eq:80a} and \eqref{eq:80b}, the remaining Bianchi
identity \eqref{eq:82} implies that $r^2\sqrt{|{}^{(2)}g|}
[G_{rr}-\frac{1}{r^2}]$ does not depend on $r$. The equation of motion
\eqref{eq:80d} then reduces to the condition
\begin{equation}
  \label{eq:89}
\hat\gamma=-\frac{l^2}{2}\bar R, 
\end{equation}
and also from the leading contribution to $K$ that
$\Theta=-\frac{l^2}{2}\bar R$.  The constraint \eqref{eq:87} then
becomes
\begin{equation}
\bar D_B \hat \gamma^{TB}_{\ A}=-\frac{l^2}{4}\d_A \bar
R.\label{eq:42}
\end{equation}

To solve this equation, one uses light-cone coordinates,
$x^\pm=\tau\pm\phi$, $2\d_\pm=\dover{}{\tau}\pm\dover{}{\phi}$ and the
explicit parameterization
$\bar\gamma_{AB}dx^Adx^B=-e^{2\varphi}dx^+dx^-$. This gives
\begin{equation}
  \label{eq:41}
  \hat \gamma=-4{l^2}e^{-2\varphi}\d_+\d_-\varphi \iff
\hat \gamma_{+-}=l^2\d_+\d_-\varphi,
\end{equation}
while the general solution to \eqref{eq:42} is 
\begin{equation}
  \label{eq:52}
  \hat\gamma_{\pm\pm}=l^2\big[\Xi_{\pm\pm}(x^\pm)
+\d^2_\pm\varphi-(\d_\pm\varphi)^2\big],
\end{equation}
with $\Xi_{\pm\pm}(x^\pm)$ 2 arbitrary functions of their
arguments. Using \eqref{eq:88},
one then gets
\begin{equation*}
  A^T_{\pm\pm}\Omega=\hat\gamma_{\pm\pm},\quad A^T_{+-}=0,\quad 
  \Omega^2=16e^{-4\varphi}\hat\gamma_{++}\hat\gamma_{--}.
\end{equation*}
In other words, one can choose $\varphi(x^+,x^-),\Xi_{\pm\pm}(x^\pm)$
as coordinates on solution space and, by expressing \eqref{eq:88} in
terms of these coordinates, we have shown that 

{\em The general solution to Einstein's equations with metrics in
  Fefferman-Graham form is given by
\begin{multline}
  g_{AB}dx^Adx^B = \Big(-e^{2\varphi} r^2 +2\hat \gamma_{+-}-
  r^{-2}e^{-2\varphi}(\hat \gamma^2_{+-}+\hat \gamma_{++}\hat
  \gamma_{--} )\Big) dx^+dx^- +\\+ \hat
  \gamma_{++}(1-r^{-2}e^{-2\varphi}\hat\gamma_{+-}) (dx^+)^2 +\hat
  \gamma_{--}(1-r^{-2}e^{-2\varphi}\hat\gamma_{+-}) (dx^-)^2,
\end{multline}
with $\hat \gamma_{AB}$ defined in equations \eqref{eq:41} and
\eqref{eq:52}. }

For instance, in these coordinates, the BTZ
black hole\cite{Banados:1992wn,Banados:1993gq} is determined by
$\varphi=0$ and
\begin{equation}
\Xi_{\pm\pm}=2G(M\pm \frac{J}{l}). 
\end{equation}

\subsection{Conformal properties of solution space}
\label{sec:conf-transf-solut}

By construction, the finite transformations generated by the spacetime
vectors \eqref{eq:FGvect} leave the Fefferman-Graham form invariant,
and furthermore transform solutions to solutions.

Using light-cone coordinates and the parametrization
$\bar\gamma_{AB}dx^Adx^B=-e^{2\varphi}dx^+dx^-$, we have 
\begin{gather*}
\left\{\begin{array}{l}
  \xi^r=-\half\psi r, 
\quad \psi=\d_+Y^++\d_-Y^-+2\d_+\varphi Y^++2\d_-\varphi Y^-, \\
  \xi^\pm=Y^\pm+\frac{l^2e^{-2\varphi}}{2r^2}\d_\mp\psi+O(r^{-4}),
\end{array}\right.
\end{gather*}
and get 
\begin{equation}
\begin{split}
  \cL_\xi g_{\pm\pm}  &\approx l^2\big[Y^\pm\d_\pm
  \Xi_{\pm\pm}+2\d_\pm Y^\pm \Xi_{\pm\pm}-\half \d^3_\pm
  Y^\pm\big]+O(r^{-2}),  \\
\cL_\xi g_{+-}   &\approx  O(r^{-2}).\label{eq:A12}
\end{split}
\end{equation}
It follows that the local conformal algebra acts on solution
space as
\begin{equation}  -\delta\Xi_{\pm\pm}=Y^\pm\d_\pm
  \Xi_{\pm\pm}+2\d_\pm Y^\pm \Xi_{\pm\pm}-\half \d^3_\pm
  Y^\pm, \label{eq:A15}
\end{equation} 
and with $\delta \varphi=0$. Note that the overall minus sign is
convential and choosen so that $\delta \Xi_{\pm\pm}\equiv
\delta_Y\Xi_{\pm\pm}$ satisfies 
$[\delta_{Y_1},\delta_{Y_2}]\Xi_{\pm\pm}=\delta_{[Y_1,Y_2]}\Xi_{\pm\pm}$. 

More generally, when considering the extension of the
conformal algebra discussed at the end of section
\bref{sec:asympt-symma}, we find that
\begin{multline*}
  \cL_\xi g_{\pm\pm}  \approx l^2\big[Y^\pm\d_\pm
  \Xi_{\pm\pm}+2\d_\pm Y^\pm \Xi_{\pm\pm}-\half \d^3_\pm
  Y^\pm+\\+\d_\pm^2\omega-2\d_\pm\varphi\d_\pm
  \omega\big]+O(r^{-2}),  
\end{multline*}
\begin{equation*}
\cL_\xi g_{+-}   \approx 2\omega (-\frac{r^2}{2}
e^{2\varphi})+l^2\d_+\d_- \omega+ O(r^{-2}), 
\end{equation*}
and thus, that the extended algebra acts on solution space as in 
 \eqref{eq:A15} with in addition $-\delta\varphi=\omega$.

\subsection{Centrally extended surface charge algebra}
\label{sec:centr-extend-surf}

Let us take 
\begin{equation}
\varphi=0\label{eq:39}.
\end{equation}
in this section. In fact, starting from a Fefferman-Graham metric
\eqref{eq:GF} with $\bar\gamma_{AB}=e^{2\varphi}\eta_{AB}$ one
can obtain such a metric with vanishing $\varphi(x^C)$ through the
finite coordinate transformation generated by $\xi^r=-\varphi r$ and
$\xi^A=-l^2\d_B\varphi\int^\infty_r\frac{dr^\prime}{r^\prime}
g^{AB}(x,r^\prime)$ since $\cL_\xi g_{rr}=0=\cL_\xi g_{rA}$ and
$\cL_\xi g_{AB}=-2\varphi g_{AB}$.

The background metric is then 
\begin{equation}
d\bar s^2 = -r^2d\tau^2 + \frac{l^2}{r^2} dr^2 + r^2
d\phi^2.
\end{equation}
Furthermore,
\begin{equation*}
Y^+=Y^\tau+Y^\phi,\ Y^-=Y^\tau-Y^\phi,\ \Lambda=\hat
\gamma_{++}+\hat \gamma_{--},\ \Sigma=\hat
\gamma_{++}-\hat \gamma_{--},\label{eq:51}
\end{equation*}
and
$g_{AB}dx^Adx^B=-r^2d\tau^2+r^2d\phi^2+h_{AB}dx^Adx^B$
with 
\begin{equation}
\begin{gathered}
  h_{\tau\tau}\approx \Lambda(x) +O(r^{-2})\approx h_{\phi\phi},\quad
  h_{\tau\phi}\approx \Sigma(x)+O(r^{-2}),\\
  \d_\tau \Lambda=\d_\phi\Sigma,\quad
  \d_\tau\Sigma=\d_\phi\Lambda.\label{eq:sol}
\end{gathered}
\end{equation}

For the surface charges, we follow \cite{Barnich:2001jy}, up to a global change of
sign, and use the expression
\begin{multline}
  \label{eq:291}
  \cQ_\xi[g-\bar g,\bar g]= \int_{\6\Sigma}\frac{\sqrt{-\bar g}}{16
    \pi G}\,(d^{n-2}x)_{\mu\nu}\, \Big[\xi^\nu\bar D^\mu h
  -\xi^\nu\bar D_\sigma h^{\mu\sigma} +\xi_\sigma\bar D^\nu
  h^{\mu\sigma}
  \\
  +\frac{1}{2}h\bar D^\nu\xi^\mu +\half
  h^{\nu\sigma}(\5D^\mu\xi_\sigma-\5D_\sigma\xi^\mu)
  -(\mu\leftrightarrow \nu)\Big].
\end{multline}
Here
\[(d^{n-k}x)_{\nu\mu}=\frac{1}{k!(n-k)!}
\epsilon_{\nu\mu\alpha_1\dots\alpha_{n-2}}
dx^{\alpha_1}\wedge\dots\wedge dx^{\alpha_{n-2}},\quad
\epsilon_{01\dots n-1}=1,\] with $n=3$ and the surface of integration
$\partial\Sigma$ is taken to be the circle at infinity. 
This gives
\begin{multline}
  \cQ_\xi[g-\bar g,\bar g]= \frac{1}{16 \pi G}\lim_{r \to \infty} \int_0^{2\pi}
  rd\phi \Big[\xi^r(\bar D^\tau h-\bar D_\sigma
  h^{\tau\sigma} +\bar D^r h^\tau_r-\bar D^\tau h^r_r)\\-\xi^\tau(\bar D^r h-\bar
  D_\sigma h^{r\sigma}-\bar D^r h^\tau_\tau+\bar D^\tau h^r_\tau)+\xi^\phi(\bar D^r
  h^{\tau}_\phi-\bar D^\tau h^{r}_\phi) +\frac{1}{2}h(\bar
  D^r\xi^\tau-\bar D^\tau\xi^r)\\+\half h^{r\sigma}(\bar D^\tau\xi_\sigma-\bar
  D_\sigma\xi^\tau)-\half  h^{\tau\sigma}(\bar D^r\xi_\sigma-\bar
  D_\sigma\xi^r)\Big].\label{eq:intch1}
\end{multline}
Using
\begin{equation}
\bar D^\tau h-\bar D_\sigma
  h^{\tau\sigma} +\bar D^r h^\tau_r-\bar D^\tau h^r_r
 =r^{-2}\bar \gamma^{AB}(\bar D_A h_{\tau B}-\bar D_\tau h_{AB})=r^{-4}(\d_\phi
 h_{\tau\phi}-\d_\tau h_{\phi\phi}),\nonumber
\end{equation} 
\begin{equation}
  \bar D^r h-\bar D_\sigma h^{r\sigma}-\bar D^r h^\tau_\tau+\bar D^\tau h^r_\tau=
 \frac{1}{l^2}(\d_r h_{\phi\phi}-\frac{1}{r} h_{\tau\tau}) ,\nonumber
\end{equation}
\begin{equation}
  \bar D^r h^\tau_\phi-\bar D^\tau h^r_\phi=
-\frac{1}{l^2}(\d_r h_{\tau \phi}-\frac{1}{r} h_{\tau \phi})\nonumber
\end{equation}
\begin{equation}
  \bar D^r\xi^\tau-\bar D^\tau\xi^r=\frac{2r}{l^2}
  Y^\tau-\frac{1}{r}\d_\tau\psi+O(r^{-3}),\nonumber
\end{equation}
\begin{equation}
\half h^{r\sigma}(\bar D^\tau\xi_\sigma-\bar
  D_\sigma\xi^\tau)-\half  h^{\tau\sigma}(\bar D^r\xi_\sigma-\bar
  D_\sigma\xi^r)=\frac{1}{rl^2}h_{\tau A} Y^A+\frac{1}{4r}
  h_{\tau}^A\d_A\psi+O(r^{-3}),\nonumber
\end{equation}
we find explicitly
\begin{multline}
\cQ_{\xi}[g-\bar g,\bar g]=\frac{1}{16\pi G l^2} \lim_{r \rightarrow
  \infty}\int_0^{2\pi}d \phi \, (2 Y^\tau h_{\phi\phi}+2
Y^{\phi}h_{\tau\phi})\\\approx 
\frac{1}{8\pi G l^2}\int_0^{2\pi}d \phi \, (
Y^\tau\Lambda+Y^\phi\Sigma)=
\frac{1}{8\pi G}\int_0^{2\pi}d \phi \, (Y^+\Xi_{++}+Y^-\Xi_{--}).
\end{multline}

The considerations of \cite{Barnich:2001jy} suggest that these charges
form a representation of the conformal algebra, or more precisely,
that 
\begin{gather}
  \cQ_{\xi_1}[\cL_{\xi_2}
  g,\bar g] \approx \cQ_{[\xi_1,\xi_2]_M}[g-\bar g,\bar g]+
K_{\xi_1 , \xi_2},\label{bracket1}\\
  K_{\xi_1 , \xi_2} = \cQ_{\xi_1}[\cL_{\xi_2} \bar g,\bar g],\quad \left[
    \xi_1,\xi_2 \right ]_M = [\xi_1,\xi_2]+\delta^g_{\xi_1}\xi_2
  -\delta^g_{\xi_2}\xi_1.\label{bracket2}
\end{gather}
An asymtotic Killing vector of the form \eqref{eq:FGvect} depends on
the metric, $\xi=\xi[x,g]$ and
$\delta^g_{\xi_1}\xi_2=\xi_2[x,\cL_{\xi_1}g]$. {}From
$\delta^g_{\xi_1}\xi_2^\tau=O(r^{-4})$ and
$\delta^g_{\xi_1}\xi_2^\phi=O(r^{-4})$, it follows that only the Lie
bracket $[\xi_1,\xi_2]$ contributes on the right hand side,
$\cQ_{[\xi_1,\xi_2]_M}[g-\bar g,\bar g]=\cQ_{[\xi_1,\xi_2]}[g-\bar
g,\bar g]$. Using \eqref{eq:A12}, \eqref{eq:sol} and
integrations by parts in $\d_\phi$ and the conformal Killing equation
for $Y^A_1,Y^A_2$ to evaluate the left hand side, one indeed finds
\begin{equation}
\begin{gathered}
  \cQ_{\xi_1}[\cL_{\xi_2} g,\bar g]  \approx
  \cQ_{[\xi_1,\xi_2]}[g-\bar g,\bar g] + K_{\xi_1 , \xi_2} ,\\
  K_{\xi_1 , \xi_2} = \frac{1}{8\pi G } \int_0^{2\pi}d \phi \,
  (\d_\phi Y^\tau_1 \partial_\phi^2 Y^\phi_2 -
  \d_\phi Y^\tau_2 \partial_\phi^2 Y^\phi_1),
\end{gathered}
\end{equation}
where $K_{\xi_1 , \xi_2}$ is a form of the well-known Brown-Henneaux
central charge.

In addition, the covariant expression for the surface charges used
above coincides on-shell with those of the Hamiltonian formalism
\cite{Barnich:2001jy,Barnich:2007bf}. In this context, it follows from
the analysis of \cite{Regge:1974zd,Brown:1986ed,Brown:1986nw} that the
surface charge is, after the Fefferman-Graham gauge fixation, the
canonical generator of the conformal transformations in the Dirac
bracket.

\section{${\rm \bf BMS_3/CFT_1}$ correspondence}
\label{sec:bms3}

We consider metrics of the form
\begin{equation}
  \label{eq:43d}
   ds^2=e^{2\beta}\frac{V}{r} du^2-2e^{2\beta}
   dudr+r^2e^{2\varphi}(d\phi-Udu)^2,
\end{equation}
or, equivalently, 
\begin{equation*}
g_{\mu\nu}=  \begin{pmatrix} e^{2\beta} V r^{-1}+r^2 e^{2\varphi} U^2 & 
  -e^{2\beta} & -r^2 e^{2\varphi} U \\
  -e^{2\beta} & 0 & 0 \\
-r^2 e^{2\varphi} U & 0 & r^2 e^{2\varphi}
\end{pmatrix}
\end{equation*}
with inverse given by 
\begin{equation*}
g^{\mu\nu}=  \begin{pmatrix} 0 & 
  -e^{-2\beta} & 0 \\
  -e^{-2\beta} & 
-\frac{V}{r} e^{-2\beta} & -U e^{-2\beta} \\
0 & -U e^{-2\beta} &  r^{-2} e^{-2\varphi}
\end{pmatrix}.
\end{equation*}
Here, $\varphi=\varphi(u,\phi)$. Three dimensional Minkowski space is
described by $\varphi=0=\beta=U$ and $V=-r$. The fall-off conditions
are taken as $\beta=O(r^{-1})$, $U=O(r^{-2})$ and $V=-2r^2\d_u
\varphi+O(r)$. In particular, $g_{uu}=-2r\d_u\varphi+O(1)$. 

\subsection{Asymptotic symmetries}
\label{sec:asympt-symm}

The transformations leaving this form of the metric invariant are
generated by vector fields such that
\begin{gather}
\cL_\xi g_{rr}=0=\cL_\xi g_{r\phi},\quad  \cL_{\xi}
g_{\phi\phi}=0,\label{eq:bms1}\\
\cL_\xi g_{ur}=O(r^{-1}),\quad \cL_\xi g_{u\phi}=O(1), \quad 
\cL_\xi g_{uu}=O(1).\label{eq:bms2}
\end{gather}
Equations \eqref{eq:bms1} imply that 
\begin{eqnarray}
  \left\{\begin{array}{l}\xi^u=f,\\
      \xi^\phi  =  Y+ I,\quad I = -e^{-2\varphi}
      \d_\phi f \,\int_r^\infty dr' \,
      {r'}^{-2} e^{2\beta}=-\frac{1}{r}e^{-2\varphi}\d_\phi f+O(r^{-2}),\\
      \xi^r =- r\big[\d_\phi \xi^\phi-\d_\phi f  U+
\xi^\phi\d_\phi \varphi +f \d_u
      \varphi \big], 
\end{array}\right.\label{eq:bms3vect}
\end{eqnarray}
with $\d_r f=0=\d_r Y$.  The first equation of \eqref{eq:bms2} then
implies that
\begin{equation}
\d_u f =f\d_u\varphi+ Y\d_\phi\varphi +\d_\phi Y\iff 
f=e^{\varphi}\big[T+\int_0^udu^\prime
e^{-\varphi}(\d_\phi Y+Y\d_\phi \varphi) \big],\label{eq:44}
\end{equation}
with $T=T(\phi)$, while the second requires $\d_u Y=0$ and thus
$Y=Y(\phi)$, which implies in turn that the last one is identically
satisfied. 

The Lie algebra $\mathfrak{bms}_3$ is determined by two arbitrary
functions $(Y,T)$ on the circle with bracket
$[(Y_1,T_1),(Y_2,T_2)]=(\hat Y,\hat T)$ determined by $\hat
Y=Y_1\d_\phi Y_2-(1\leftrightarrow 2)$ and $\hat T= Y_1\d_\phi T_2
+T_1\d_\phi Y_2-(1\leftrightarrow 2)$. Let $\Im=S^1\times {\mathbb R}$
with coordinates $u,\phi$ and consider the vector fields $\bar\xi=f
\dover{}{u}+ Y\dover{}{\phi}$ with $f$ as in \eqref{eq:44} and
$Y=Y(\phi)$. By direct computation, it follows that these vector
fields equipped with the commutator bracket provide a faithful
representation of $\mathfrak{bms}_3$. Furthermore : 

{\em The spacetime vectors \eqref{eq:bms3vect}, with $f$ given in 
\eqref{eq:44} and $Y=Y(\phi)$ form a faithful
representation of the $\mathfrak{bms}_3$ Lie algebra on an
asymptotically flat spacetime of the form \eqref{eq:43d} when equipped
with the modified bracket $[\cdot,\cdot]_M$. }

Indeed, for the $u$ component, there is no modification due to the
change in the metric and the result follows by direct computation.  As
a consequence, $\hat f=[\xi_1,\xi_2]^u_{(M)}$ corresponds to $f$ in
\eqref{eq:44} with $T$ replaced by $\hat T$ and $Y$ by $\hat Y$. By
evaluating $\cL_\xi g^{\mu\nu}$, we find
\begin{gather}
\left\{\begin{array}{l}
  \label{eq:10}
\delta_\xi\varphi=0,\\
  \delta_\xi\beta =\xi^\alpha\d_\alpha\beta+\half
  \big[\d_u f+\d_r\xi^r+\d_\phi f U],\\
\delta_\xi U =\xi^\alpha \partial_\alpha U + U \big[\d_u f+ \partial_\phi f
U-\d_\phi \xi^\phi\big]
- \partial_u \xi^\phi-\d_r \xi^\phi\frac{V}{r}
+\partial_\phi \xi^r\frac{e^{2(\beta-\varphi)}}{r^2}.
\end{array}
\right.
\end{gather}
It follows that 
\begin{gather}
  \label{eq:11}
\left\{\begin{array}{l}
  \delta^g_{\xi_1}\xi^\phi_2=-e^{-2\varphi}\d_\phi
  f_2\int^\infty_r\frac{dr^\prime}{{r^{\prime}}^2}
    e^{2\beta}2\delta_{\xi_1}\beta,\\
\delta^g_{\xi_1}\xi^r_2=-r\big[\d_\phi(\delta^g_{\xi_1}\xi^\phi_2)
+(\delta^g_{\xi_1}\xi^\phi_2)\d_\phi
\varphi-\d_\phi f_2 \delta_{\xi_1} U\big].
\end{array}
\right.
\end{gather}
We also have $\lim_{r\to\infty}[\xi_1,\xi_2]^\phi_M=\hat Y$. Using
$\d_r\xi^\phi=\frac{e^{2(\beta-\varphi)}}{r^2}\d_\phi f$,
\eqref{eq:44} and the expression of $\xi^r$ in \eqref{eq:bms3vect}, it
follows by a straightforward computation that
$\d_r([\xi_1,\xi_2]^\phi_M)=\frac{e^{2(\beta-\varphi)}}{r^2}\d_\phi\hat
f$, which gives the result for the $\phi$ component.  Finally, for the
$r$ component, we need the relation
\[\d_r(\frac{\xi^r}{r})=-\d_r\big(\d_\phi\xi^\phi
+\xi^\phi\d_\phi f\d_\phi\varphi-\d_\phi
fU\big).\] We then have
$\lim_{r\to\infty}\frac{[\xi_1,\xi_2]^r_M}{r}=-\d_\phi\hat Y-\hat Y
\d_\phi\varphi-\hat f\d_u\varphi$, while direct computation shows that
$\d_r(\frac{[\xi_1,\xi_2]^r_M}{r})=-\d_r\big(
\d_\phi([\xi_1,\xi_2]^\phi_M)
-\d_\phi([\xi_1,\xi_2]^u_M)U+[\xi_1,\xi_2]^\phi_M\d_\phi\varphi\big)$, 
which gives the result for the
$r$ component.

More generally, one can also consider the transformations that leave
the form of the metric \eqref{eq:43d} invariant up to a rescaling of
$\varphi$ by $\omega(u,\phi)$. They are generated by spacetime vectors
satisfying
\begin{gather}
\cL_\xi g_{rr}=0=\cL_\xi g_{r\phi},\quad  \cL_{\xi}
g_{\phi\phi}=2\omega g_{\phi\phi},\label{eq:bms1a}\\
\cL_\xi g_{ur}=O(r^{-1}),\quad \cL_\xi g_{u\phi}=O(1), \quad 
\cL_\xi g_{uu}=-2r\d_u\omega +O(1).\label{eq:bms2a}
\end{gather}
Equations \eqref{eq:bms1a}, \eqref{eq:bms2a} then imply that the
vectors are given by \eqref{eq:bms3vect}, \eqref{eq:44} with the
replacement $\d_\phi Y\to \d_\phi Y-\omega$.

With this replacement, the vector fields $\bar\xi=f \dover{}{u}+
Y\dover{}{\phi}$ on $\Im=S^1\times {\mathbb R}$ equipped with the
modified bracket provide a faithful representation of the extension of
$\mathfrak{bms}_3$ defined by elements $(Y,T,\omega)$ and bracket
$[(Y_1,T_1,\omega_1),(Y_2,T_2,\omega_2)]=(\hat Y,\hat T,\hat\omega)$,
with $\hat Y,\hat T$ as before and $\hat \omega=0$.

Indeed, the result is obvious for the $\phi$ component. Furthermore,
\[\delta^g_{\bar \xi_1} f_2= \omega_1 f_2+e^{\varphi}\int_0^udu^\prime
e^{-\varphi}[-\omega_1(\d_\phi Y_2-\omega_2+Y_2\d_\phi\varphi)+
Y_2\d_\phi \omega_1].\] At $u=0$, we get
$[\bar\xi_1,\bar\xi_2]^u_M|_{u=0}=e^\varphi|_{u=0} \hat T$, while
direct computation shows that $\d_u ([\bar\xi_1,\bar\xi_2]^u_M)=\hat
f\d_u\varphi +\hat Y\d_\phi\varphi+\d_\phi\hat Y$, as it
should. 

Following the same reasoning as before, one can then also show that
the spacetime vectors \eqref{eq:bms3vect} with the replacement
discussed above and equipped with the modified Lie bracket provide a
faithful representation of the extended $\mathfrak{bms}_3$ algebra. 

Indeed, we have $\xi=\bar\xi +I\dover{}{\phi}+\xi^r\dover{}{r}$.  Furthermore,
$[\xi_1,\xi_2]_M^u=[\bar\xi_1,\bar\xi_2]^u_M=\hat f$ as it should.  In
the extended case, the variations of $\beta,U$ are still given by
\eqref{eq:10}.  We then have
$\lim_{r\to\infty}[\xi_1,\xi_2]_M^\phi=\hat Y$ and find, after some
computations, $\d_r
([\xi_1,\xi_2]_M^\phi)=\frac{e^{2(\beta-\varphi)}}{r^2} \d_\phi\hat
f$, giving the result for the $\phi$ component. Finally, we have
$\lim_{r\to\infty}\frac{[\xi_1,\xi_2]^r_M}{r}=-\d_\phi\hat Y-\hat Y
\d_\phi\varphi-\hat f\d_u\varphi$, while direct computation shows that
$\d_r(\frac{[\xi_1,\xi_2]^r_M}{r})=-\d_r\big(
\d_\phi([\xi_1,\xi_2]^\phi_M)
-\d_\phi([\xi_1,\xi_2]^u_M)U+[\xi_1,\xi_2]^\phi_M\d_\phi\varphi
\big)$, which gives the result for the $r$ component.

\subsection{Solution space}
\label{sec:solution-space}

Following \cite{Tamburino:1966}, the equations of motion are organized
in terms of the Einstein tensor $G_{\alpha\beta}=R_{\alpha\beta}-\half
g_{\alpha\beta} R$ as
\begin{gather}
  G_{r\alpha}=0,  \qquad
G_{AB}-\half g_{AB} g^{CD}G_{CD}=0,  \label{eq:571b}\\
G_{uu}=0=G_{uA},  \label{eq:571c}\\
g^{CD}G_{CD}=0,  \label{eq:571d} 
\end{gather}
and the Bianchi identities are written as 
\begin{equation}
\label{eq:561}
 0= 2\sqrt{-g}{G_\alpha^\beta}_{;\beta}=2(\sqrt{-g}G_\alpha^\beta)_{,\beta}+
\sqrt{-g}G_{\beta\gamma}{g^{\beta\gamma}}_{,\alpha}. 
\end{equation}

For a metric of the form \eqref{eq:43d}, we have 
\begin{equation*}
\begin{gathered}
\Gamma^\lambda_{rr}=\delta^\lambda_r2\beta_{,r},\quad
\Gamma^{u}_{\lambda r}=0,\quad
\Gamma^r_{\phi r}=\beta_{,\phi }+ n ,\quad \Gamma^\phi _{\phi r}=\frac{1}{r} ,\\
\Gamma^u_{\phi \phi }= e^{-2\beta+2 \varphi} r,\quad
\Gamma^\phi _{\phi \phi}= e^{-2\beta+2 \varphi} U r+\d_\phi\varphi, \\ 
\Gamma^\phi _{ur}=
-\frac{1}{r} U +r^{-2}e^{2\beta- 2 \varphi}(\d_\phi \beta-n ),\quad 
\Gamma^u_{u\phi }=\beta_{,\phi }-n - e^{-2\beta+2 \varphi} rU ,\\
\Gamma^r_{ur}=-\half (\d_r+2\beta_{,r})\frac{V}{r}-(\beta_{,\phi }+n )
U ,\quad 
\Gamma^\phi _{\phi u}=\d_u \varphi   +U (\beta_{,\phi }-n )- e^{-2\beta+2\varphi}rU^2,\\
\Gamma^{u}_{uu}=2\beta_{,u}+\half(\d_r+2\beta_{,r})
\frac{V}{r}+2U  n +e^{-2\beta+2\varphi}r U^2 ,\\
\Gamma^r_{\phi \phi }= e^{-2\beta+2 \varphi}(r^2\d_\phi  U +r^2
\d_\phi \varphi  U 
+r^2\d_u \varphi+V ),\\
\Gamma^r_{u\phi }=-\frac{V_{,\phi }}{2r}-n \frac{V}{r}-
e^{-2\beta+2\varphi}U [r^2\d_\phi  U +r^2 \d_\phi \varphi  U -r^2\d_u \varphi+V ],\\
\Gamma^\phi _{uu}=2U \beta_{,u}+\half
U (\d_r+2\beta_{,r})\frac{V}{r}+2U^2  n +re^{-2\beta+2 \varphi}U^3
-U_{,u}-2\d_u \varphi  U\\-\half
e^{2\beta-2 \varphi}r^{-2}(\d_\phi +2\d_\phi \beta)\frac{V}{r}-U (\d_\phi +\d_\phi \varphi) U,\\
\Gamma^r _{uu}=- \half (\d_u - 2 \d_u \beta)\frac{V}{r} + \half
\frac{V}{r}(\d_r + 2 \d_r \beta) \frac{V}{r} + V r e^{2 \varphi - 2
  \beta} U (\d_r + \frac{1}{r})U + r^2 e^{-2\beta + 2 \varphi} U^2
\d_u \varphi \\
+ \half U (\d_\phi + 2 \d_\phi \beta)\frac{V}{r} 
+ \half r^2 e^{-2 \beta + 2 \varphi} U (\d_\phi + 2 \d_\phi \varphi) U,
\end{gathered}
\end{equation*}
where the notation $n=\half r^2 e^{2\varphi-2\beta} \d_r U$ has been
used.

We start with $G_{rr}=0$. From
\begin{equation*}
G_{rr}=R_{rr}=\frac{2}{r} \d_r \beta,
\end{equation*}
we find $\beta=0$ by taking the fall-off conditions into
account. From
\begin{equation*}
  G_{r\phi}=R_{r\phi}=(\d_r +\frac{1}{r}) n + \frac{1}{r}\d_\phi \beta,
\end{equation*}
we then obtain, by using the previous result, that $n=\frac{N}{r}$
where the integration constant $N=N(u,\phi)$.  Using the definition of
$n$, we get $U=-r^{-2} e^{-2\varphi} N$.  From
$G_{ru}=-g^{\phi\phi}R_{\phi\phi}$ and
\begin{multline*}
R_{\phi\phi}= e^{-2\beta+2\varphi} 
\left( (\d_r - \frac{1}{r})V + 2r \d_u \varphi + 2r(\d_\phi+\d_\phi \varphi) U \right)\\
-2 \d_\phi^2 \beta + 2 \d_\phi \beta \d_\phi \varphi 
- 2 (\d_\phi \beta - n)^2 -2 \d_\phi\varphi n + 2 \d_\phi n\\
  = e^{2\varphi} \left( (\d_r - \frac{1}{r})V + 2r \d_u \varphi\right)- 2 r^{-2 } N^2,
\end{multline*}
we get $\d_r (\frac{V}{r}) = - 2 \d_u \varphi + 2 r^{-3} e^{-2\varphi} N^2$
and then 
\begin{equation*}
V=-2r^2 \d_u \varphi + r M - r^{-1} e^{-2\varphi}N^2,
\end{equation*}
for an additional integration constant $M=M(u,\phi)$. 

When $G_{rr}=G_{r\phi}=G_{ru}=0$, the Bianchi identity \eqref{eq:561}
for $\alpha=r$ implies that $G_{\phi\phi}=0$. This implies in turn
that $R=0$. The Bianchi identity for $\alpha=\phi$ then gives $\d_r (r
G_{u\phi})=0$. When $G_{u\phi}=0$, the Bianchi identity for $\alpha=u$
gives $\d_r (r G_{uu})=0$. To solve the remaining equations of motion,
there thus remain only the constraints
\begin{equation*}
  \lim_{r \rightarrow \infty} r R_{u\phi} =0,\qquad \qquad 
\lim_{r \rightarrow \infty} r R_{uu} =0.
\end{equation*}
to be fulfilled. From 
\begin{equation*}
R_{u\phi} = \frac{1}{r} \left( -(\d_u + \d_u \varphi) N + 
\half \d_\phi M\right) + O(r^{-2}), 
\end{equation*}
we get 
\begin{equation*}
N=e^{-\varphi}\, \Xi(\phi)+ e^{-\varphi} \int^u_{u_0} d\tilde u \, e^\varphi \half \d_\phi M.
\end{equation*}
while 
\begin{equation*}
R_{uu} = \frac{1}{r} \left( -\half(\d_u + 2\d_u \varphi) M + 
e^{-2\varphi}\d_u(\d_\phi^2 \varphi-\half (\d_\phi \varphi)^2)\right) + O(r^{-2})
\end{equation*}
implies
\begin{equation*}
M = e^{-2\varphi}[\Theta (\phi) -(\d_\phi
  \varphi)^2+2\d_\phi^2 \varphi]. 
\end{equation*}
We thus have shown:

{\em For metrics of the form \eqref{eq:43d} with $\lim_{r \rightarrow
    \infty}\beta=0$, the general solution to the equations of motions
  is given by
\begin{equation}
\begin{gathered}
  \label{eq:44b}
  ds^2=s_{uu}du^2-2dudr+2s_{u\phi}dud\phi+r^2e^{2\varphi}d\phi^2,\\
s_{uu}=
e^{-2\varphi}\big[\Theta-(\d_\phi\varphi)^2+2\d^2_\phi\varphi\big]-2r\d_u
\varphi,\\
s_{u\phi}=e^{-\varphi}\Big[\Xi+\int^u_{u_0} d\tilde u
e^{-\varphi}\big[\half\d_\phi\Theta-\d_\phi\varphi[\Theta-
(\d_\phi\varphi)^2+3\d^2_\phi\varphi]+\d^3_\phi
\varphi\big]\Big],
\end{gathered}
\end{equation}
where $\Theta=\Theta(\phi)$ and $\Xi=\Xi(\phi)$ are arbitrary
functions. 
}

\subsection{Conformal properties of solution space}
\label{sec:conf-transf-solutbis}

By computing $\cL_\xi s_{\mu\nu}$, we find that the asymptotic
symmetry algebra $\mathfrak{bms}_3$ acts on solution space according
to 
\begin{equation}
\begin{split}
  -\delta\, \Theta & = Y\d_\phi \Theta+2 \d_\phi Y \Theta -
  2 \partial^3_\phi Y, \label{eq:B11}\\
  -\delta\, \Xi & = Y \partial_\phi\Xi+ 2 \d_\phi Y \Xi  +\half
  T\d_\phi\Theta+ \d_\phi T \Theta-\d^3_\phi
  T, \\
 -\delta\,\varphi &=0. 
\end{split}
\end{equation}
For the extended algebra, the first two relations are unchanged, while
$-\delta\varphi=\omega$. 

\subsection{Centrally extended surface charge algebra}
\label{sec:centr-extend-charge}

Let us again take $ \varphi=0$ in this section.  For the surface
charges computed at the circle at infinity $u=cte, r=cte\to\infty$,
one starts again from \eqref{eq:291}. The background line element,
which is used to raise and lower indices, is
\begin{equation}
  \label{eq:20a}
  d\bar s^2=-du^2-2dudr+r^2 d\phi^2, 
\end{equation}
This gives
\begin{multline}
  \cQ_\xi[g-\bar g,\bar g]= \frac{1}{16 \pi G}\lim_{r \to \infty} \int_0^{2\pi}
  rd\phi \Big[\xi^r(\bar D^u h-\bar D_\sigma
  h^{u\sigma} +\bar D^r h^u_r-\bar D^u h^r_r)\\-\xi^u(\bar D^r h-\bar
  D_\sigma h^{r\sigma}-\bar D^r h^u_u+\bar D^u h^r_u)+\xi^\phi(\bar D^r
  h^{u}_\phi-\bar D^u h^{r}_\phi) +\frac{1}{2}h(\bar
  D^r\xi^u-\bar D^u\xi^r)\\+\half h^{r\sigma}(\bar D^u\xi_\sigma-\bar
  D_\sigma\xi^u)-\half  h^{u\sigma}(\bar D^r\xi_\sigma-\bar
  D_\sigma\xi^r)\Big].\label{eq:intch2}
\end{multline}
Using
\begin{equation}
\bar D^u h-\bar D_\sigma
  h^{u\sigma} +\bar D^r h^u_r-\bar D^u h^r_r
=-\frac{1}{r} h_{ur},\nonumber
\end{equation} 
\begin{equation}
  \bar D^r h-\bar D_\sigma h^{r\sigma}-\bar D^r h^u_u+\bar D^u h^r_u
 =-\frac{1}{r}h_{uu} +\frac{2}{r} h_{ur}+ \frac{1}{r^{2}}\d_\phi h_{u\phi},\nonumber
\end{equation}
\begin{equation}
  \bar D^r h^u_\phi-\bar D^u h^r_\phi=(\frac{1}{r}-\d_r) h_{u\phi},\nonumber
\end{equation}
\begin{equation}
  \bar D^r\xi^u-\bar D^u\xi^r=-2\d_\phi Y+O(1),\nonumber
\end{equation}
\begin{equation}
\half h^{r\sigma}(\bar D^u\xi_\sigma-\bar
  D_\sigma\xi^u)-\half  h^{u\sigma}(\bar D^r\xi_\sigma-\bar
  D_\sigma\xi^r)=-2\d_\phi Y h_{ur}+\frac{1}{r}h_{u\phi}Y+O(r^{-2}), \nonumber
\end{equation}
we get
\begin{multline}
   \cQ_\xi[g-\bar g,\bar g]=\frac{1}{16 \pi G}\lim_{r \to \infty}
   \int_0^{2\pi}d\phi\, 
\Big[ (r h_{ur}+u h_{uu})\d_\phi Y+ h_{uu}T +2 h_{u\phi}
Y\Big]\\
 \approx\frac{1}{16 \pi G}
   \int_0^{2\pi}d\phi\, (\Theta T +2\Xi Y).  \label{eq:36b}
\end{multline}
The surface charges thus provide the inner product that turns the
linear spaces of solutions and asymptotic symmetries into dual
spaces. It follows that the solutions that we have constructed are all
non-trivial as different solutions carry different charges. 

{}From $\delta^g_{\xi_1}\xi_2^u=0$,
$\delta^g_{\xi_1}\xi_2^u=O(r^{-2})$ and
$\delta^g_{\xi_1}\xi_2^r=O(r^{-1})$, it follows that only the Lie
bracket $[\xi_1,\xi_2]$ contributes on the right hand side of
\eqref{bracket1}-\eqref{bracket2}, $\cQ_{[\xi_1,\xi_2]_M}[g-\bar
g,\bar g]=\cQ_{[\xi_1,\xi_2]}[g-\bar g,\bar g]$. Using \eqref{eq:36b},
\eqref{eq:B11} and integrations by parts in $\d_\phi$ to evaluate the
left hand side, one indeed finds
\begin{align}
  \cQ_{\xi_1}[\cL_{\xi_2} g,\bar g] & \approx
  \cQ_{[\xi_1,\xi_2]}[g-\bar g,\bar g] + K_{\xi_1 , \xi_2} ,\\
  K_{\xi_1 , \xi_2} &= \frac{1}{8\pi G } \int_0^{2\pi}d \phi \,
  \Big[\d_\phi Y_1(T_2+\d^2_\phi T_2)-\d_\phi Y_2(T_1+\d^2_\phi T_1)\Big].
\end{align}
where $K_{\xi_1 , \xi_2}$ is the central charge\footnote{Due to a
  change of conventions, some sign errors remain in section 2 of the
  published version of \cite{Barnich:2006avcorr}. See the latest arxiv
  version for corrections.}.

\section{${\rm \bf BMS_4/CFT_2}$ correspondence} 
\label{sec:asympt-flat-spac}

\subsection{Asymptotically flat $4$-d spacetimes at null infinity}
\label{sec:asympt-flat-4}

Let $x^0=u,x^1=r,x^2=\theta,x^3=\phi$ and $A,B,\dots=2,3$. Following
mostly Sachs \cite{Sachs2} up to notation, the metric $g_{\mu\nu}$ of an
asymptotically flat spacetime can be written in the form
\begin{equation}
  \label{eq:2}
  ds^2=e^{2\beta}\frac{V}{r} du^2-2e^{2\beta}dudr+
g_{AB}(dx^A-U^Adu)(dx^B-U^Bdu)
\end{equation}
where $\beta,V,U^A,g_{AB} (\text{det}\, g_{AB})^{-1/2}$ are $6$
functions of the coordinates, with $\text{det}\, g_{AB}=r^4
b(u,\theta,\phi)$ for some fixed function $b(u,\theta,\phi)$.  The
inverse to the metric
\begin{equation*}
g_{\mu\nu}=  \begin{pmatrix} e^{2\beta} \frac{V}{r}+g_{CD}U^CU^D & 
  -e^{2\beta} & -g_{BC} U^C \\
  -e^{2\beta} & 0 & 0 \\
-g_{AC} U^C & 0 & g_{AB} 
\end{pmatrix}
\end{equation*}
is given by 
\begin{equation*}
g^{\mu\nu}=  \begin{pmatrix} 0 & 
  -e^{-2\beta} & 0 \\
  -e^{-2\beta} & 
-\frac{V}{r} e^{-2\beta} & -U^B e^{-2\beta} \\
0 & -U^A e^{-2\beta} & g^{AB}
\end{pmatrix}.
\end{equation*}
The fall-off conditions are 
\begin{equation}
  \label{eq:3a}
g_{AB}dx^Adx^B=r^2\bar \gamma_{AB}dx^Adx^B+O(r),
\end{equation}
Sachs chooses $\bar\gamma_{AB}={}_0\gamma_{AB}$ to be the metric on
the unit $2$ sphere, ${}_0\gamma_{AB}dx^Adx^B=d\theta^2+\sin^2\theta
d\phi^2$ and $b=\sin^2\theta$, but the geometrical analysis by Penrose
\cite{PhysRevLett.10.66} suggests to be somewhat more general and use
a metric that is conformal to the latter, for instance,
$\bar\gamma_{AB}dx^Adx^B=e^{2\varphi}(d\theta^2+\sin^2\theta
d\phi^2)$, with $\varphi=\varphi(u,x^A)$. We will choose the
determinant condition more generally to be
$b(u,x^A)=\text{det}\bar\gamma_{AB}$. In particular, in the above
example, on which we now focus, $b=e^{4\varphi}\sin^2\theta$.

The rest of the fall-off conditions are given by
\begin{equation}
\label{eq:3b}
\beta=O(r^{-2}),\quad V/r=-2r\dot\varphi-e^{-2\varphi}+
  \bar\Delta\varphi+O(r^{-1}), 
\quad U^A=O(r^{-2}).
\end{equation}
Here, a dot denotes the derivative with respect to $u$, $\bar D_A$
denotes the covariant derivative with respect to $\bar
\gamma_{AB}$. We denote by $\bar \Gamma^A_{BC}$ the associated
Christoffel symbols and by $\bar\Delta$ the associated
Laplacian. Similiarily, ${}_0 D_A,{}_0\Gamma^A_{BC},{}_0\Delta$
correspond to ${}_0\gamma_{AB}$.  Note that $g^{AB}g_{BC}=\delta^A_C$
and that the condition on the determinant implies
\begin{gather}
  \label{eq:27}
\left\{\begin{array}{l}
  g^{AB}\d_rg_{AB}=4 r^{-1},\\
g^{AB}\d_u g_{AB}=\bar\gamma^{AB}\d_u\bar\gamma_{AB}=4\dot\varphi, \\
g^{AB}\d_C g_{AB}=\bar
\gamma ^{AB}{}\d_C\bar
\gamma_{AB}={}_0\gamma^{AB}\d_C\,{}_0\gamma_{AB}+4\d_C\varphi,
\end{array}\right.
\end{gather}
where $\bar\gamma^{AB}\bar\gamma_{BC}=\delta^A_C={}_0\gamma^{AB}{}_0
\gamma_{BC}$.  In terms of the metric and its inverse, the fall-off
conditions read
\begin{gather*}
\left\{\begin{array}{l}
g_{uu}=-2r\dot\varphi-e^{-2\varphi}+\bar\Delta\varphi+O(r^{-1}),\ 
g_{ur}=-1+O(r^{-2}),\
g_{uA}=O(1), \\
g_{rr}=0=g_{rA},\quad 
g_{AB}=r^2\bar\gamma_{AB}+O(r),\\
g^{ur}=-1+O(r^{-2}),\quad g^{uu}=0=g^{uA},\\
g^{rr}=2r\dot\varphi+e^{-2\varphi}-\bar\Delta\varphi+O(r^{-1}),\ g^{rA}=O(r^{-2}),\
g^{AB}=r^{-2}\bar\gamma^{AB}+O(r^{-3}).
\end{array}\right.
\end{gather*}

With the choice $\varphi=0$, Sachs studies the vector fields that
leave invariant this form of the metric with these fall-off
conditions. More precisely, he finds the general
solution to the equations
\begin{gather}
  \label{eq:4a}
  \cL_\xi g_{rr}=0,\quad \cL_\xi g_{rA}=0,\quad \cL_\xi g_{AB}
  g^{AB}=0,\\
\cL_\xi
g_{ur}=O(r^{-2}),\quad \cL_\xi g_{uA}=O(1),\quad  \cL_\xi
g_{AB}=O(r),\quad 
\cL_\xi g_{uu}=O(r^{-1}). \label{eq:4b}
\end{gather}
For arbitrary $\varphi$, the general solution to \eqref{eq:4a} is given by 
\begin{gather}
  \label{eq:26}
\left\{\begin{array}{l}
  \xi^u=f,\\
\xi^A=Y^A+I^A, \quad  I^A=- f_{,B} \int_r^\infty dr^\prime(
e^{2\beta} g^{AB}),\\
\xi^r=-\half r (\psi+\chi-f_{,B}U^B+2f\d_u\varphi),
\end{array}\right.
\end{gather}
with $\d_r f=0=\d_r Y^A$ and where $ \psi= \bar D_A Y^A$, $\chi= \bar
D_A I^A$.  This gives the expansions
\begin{gather}
  \label{eq:14}
  \left\{\begin{array}{l} \xi^u=f,\quad
      \xi^A=Y^A-r^{-1} f_{,B}\bar\gamma^{BA}+O(r^{-2}),\\
      \xi^r=-r(f\dot\varphi+\frac{1}{2}\psi)+\half \bar\Delta f +O(r^{-1}).
\end{array}\right.
\end{gather}
The first equation of \eqref{eq:4b} then implies that
\begin{equation}
\dot f =f\dot\varphi+\half \psi\iff f=e^{\varphi}\big[T+
 \half\int_0^udu^\prime
e^{-\varphi}\psi\big],\label{eq:44ter}
\end{equation}
with $T=T(\theta,\phi)$, while the second requires $\d_u Y^A=0$ and
thus $Y^A=Y^A(\theta,\phi)$. The third one implies that $Y^A$ is a
conformal Killing vector of $\bar\gamma_{AB}$ and thus also of
${}_0\gamma_{AB}$. The last equation of \eqref{eq:4b} is then
satisfied without additional conditions. For the computation, one uses
that $\bar\Delta =e^{-2\varphi}{}_0\Delta $ and $\psi={}_0 \psi +2 Y^A\d_A
\varphi$, with ${}_0\psi={}_0D_A Y^A$ and the following properties
of conformal Killing vectors $Y^A$ on the unit $2$-sphere,
\begin{equation}
  \label{eq:35}
  2{}_0 D_B{}_0 D_C Y_A={}_0 \gamma_{CA}{}_0 D_B{}_0\psi+{}_0 \gamma_{AB}
{}_0 D_C{}_0\psi-{}_0 \gamma_{BC}{}_0 D_A{}_0 \psi+2
  Y_C {}_0 \gamma_{BA}-2 Y_A {}_0 \gamma_{BC},  
\end{equation}
where the indices on $Y^A$ are lowered with $\bar\gamma_{AB}$.  This
implies in particular ${}_0 \Delta Y^A=-Y^A$ and also that ${}_0
\Delta {}_0\psi =-2{}_0\psi$.

\subsection{ $\mathfrak{bms}_4$ Lie algebra}
\label{sec:mathfr-algebra}

By definition, the algebra $\mathfrak{bms}_4$ is the semi-direct sum
of the Lie algebra of conformal Killing vectors $Y^A\dover{}{x^A}$ of the unit
$2$-sphere with the abelian ideal consisting of functions
$T(x^A)$ on the $2$-sphere. The bracket is defined through 
\begin{equation}
\label{eq:32}
\begin{split}
(\hat Y,\hat T)&=[(Y_1,T_1),(Y_2,T_2)], \\
\hat Y^A&= Y^B_1\d_B
Y^A_2-Y^B_1\d_B Y^A_2,\\
\hat T&=Y^A_1\d_A
  T_2-Y^A_2\d_A T_1 +\half (T_1\,{}_0\psi_2-T_2\,{}_0\psi_1).
\end{split}
\end{equation}

Let $\Im={\mathbb R}\times S^2$ with coordinates $u,\theta,\phi$. On
$\Im$, consider the scalar field $\varphi$ and the vectors fields
$\bar\xi(\varphi,T,Y)=f\dover{}{u} +Y^A\dover{}{x^A}$, with $f$ given
in \eqref{eq:44ter} and $Y^A$ an $u$-independent conformal Killing
vector of $S^2$. It is straightforward to check that these vector
fields form a faithful representation of $\mathfrak{bms}_4$ for the
standard Lie bracket.

Consider then the modified Lie bracket
\begin{equation}
  \label{eq:43}
  [\xi_1,\xi_2]_M=[\xi_1,\xi_2]-\delta^g_{
    \xi_1}\xi_2+\delta^g_{ \xi_2}\xi_1,
\end{equation}
where $\delta^g_{\xi_1}\xi_2$ denotes the variation in $\xi_2$
under the variation of the metric induced by $\xi_1$, $\delta^g_{
  \xi_1}g_{\mu\nu}=\cL_{\xi_1}g_{\mu\nu}$, 

{\em Spacetime vectors $\xi$ of the form \eqref{eq:26}, with
  $Y^A(x^B)$ a conformal Killing vectors of the $2$-sphere and
  $f(u,x^B)$ satisfying \eqref{eq:44ter} provide a faithful
  representation of $\mathfrak{bms}_4$ when equipped with the modified
  Lie bracket $[\cdot,\cdot]_M$.}

Indeed, for the $u$ component, there is no modification due to the
change in the metric and the result follows by direct computation:
$[\xi_1,\xi_2]^u_{(M)}=\hat f$, where $\hat f$ corresponds to $f$ in
\eqref{eq:44ter} with $T$ replaced by $\hat T$ and $Y$ by $\hat Y$. By
evaluating $\cL_\xi g^{\mu\nu}$, we find
\begin{gather}
\left\{\begin{array}{l}
  \label{eq:10bis}
 \delta_\xi\varphi=0,\\
  \delta_\xi\beta=\xi^\alpha\d_\alpha\beta+\half \big[\d_u f
  +\d_r\xi^r+\d_A f U^A\big],\\
  \delta_\xi U^A = \xi^\alpha \partial_\alpha U^A + U^A (\partial_u f
  + \partial_B f U^B) -\d_B\xi^A U^B\\\hspace{1cm} - \partial_u \xi^A
  - \partial_r \xi^A \frac{V}{r} + \partial_B \xi^r
  g^{AB}e^{2\beta}.
\end{array}
\right.
\end{gather}
It follows that 
\begin{gather*}
\left\{\begin{array}{l}
  \delta^g_{\xi_1}\xi^A_2=-\d_B
  f_2\int^\infty_r dr^\prime
    e^{2\beta}(2\delta_{\xi_1}\beta g^{AB}+\cL_\xi g^{AB}),\\
\delta^g_{\xi_1}\xi^r_2=-\half r\big[\bar D_A (\delta^g_{\xi_1}\xi^A_2)
-\d_A f_2 \delta_{\xi_1} U^A\big].
\end{array}
\right.
\end{gather*}
We have $\lim_{r\to\infty}[\xi_1,\xi_2]^A_M=\hat Y^A$ and, using
$\d_r\xi^A=g^{AB}e^{2\beta}\d_B f$, \eqref{eq:44ter} together with the expression
of $\xi^r$ in \eqref{eq:14}, it follows by a straightforward
computation that $\d_r([\xi_1,\xi_2]^A_M)=g^{AB} e^{2\beta}\d_B\hat
f$, which gives the result for the $A$ components.  Finally, for the
$r$ component, we need
\[\d_r(\frac{\xi^r}{r})=-\half\big(\d_r\chi-\d_B f \d_r U\big).\]
We then find $\lim_{r\to\infty}\frac{[\xi_1,\xi_2]^r_M}{r}=-\half
(\hat \psi+2\hat f \d_u\varphi)$, where $\hat \psi$ corresponds to
$\psi$ with $Y^A$ replaced by $\hat Y^A$, while
$\d_r(\frac{[\xi_1,\xi_2]^r_M}{r})=-\half (\d_r \hat \chi-\d_B\hat
f\d_r U^B)$, where $\hat \chi$ corresponds to $\chi$ with $f$ replaced
by $\hat f$. This gives the result for the $r$ component and concludes
the proof.

In terms of the standard complex coordinates $\zeta=
e^{i\phi}\cot{\frac{\theta}{2}}$, the metric on the sphere is
conformally flat, 
  \begin{equation}
    \label{eq:6}
    d\theta^2+\sin^2\theta
    d\phi^2=P^{-2}d\zeta
    d\bar\zeta,\quad P(\zeta,\bar\zeta)=\half(1+\zeta\bar\zeta), 
  \end{equation}
and, since conformal Killing vectors are invariant under conformal
rescalings of the metric, the conformal Killing vectors of the unit
sphere are the same as the conformal Killing vectors of the
Riemann sphere. 

Depending on the space of functions under consideration, there are
then basically two options which define what is actually meant by
$\mathfrak{bms}_4$.

The first choice consists in restricting oneself to globally
well-defined transformations on the unit or, equivalently, the Riemann
sphere. This singles out the global conformal transformations, also
called projective transformations, and the associated group is
isomorphic to $SL(2,\mathbb C)/\mathbb Z_2$, which is itself
isomorphic to the proper, orthochronous Lorentz group. Associated with
this choice, the functions $T(\theta,\phi)$, which are the generators
of the so-called supertranslations, have been expanded into spherical
harmonics. This choice has been adopted in the original work by Bondi,
van der Burg, Metzner and Sachs and followed ever since, most notably
in the work of Penrose and Newman-Penrose
\cite{PhysRevLett.10.66,newman:863}, where spin-weighted spherical
harmonics and the associated ``edth'' operator have made their
appearance. After attempts to cut this group down to the standard
Poincar\'e group, it has been taken seriously as an invariance group
of asymptotically flat spacetimes. Its consequences have been
investigated, but we believe that it is fair to say that this version
of the BMS group has had only a limited amount of success.

The second choice that we would like to advocate here is motivated by
exactly the same considerations that are at the origin of the
breakthrough in two dimensional conformal quantum field theories
\cite{Belavin:1984vu}. It consists in focusing on local properties and
allowing the set of all, not necessarily invertible holomorphic
mappings. In this case, Laurent series on the Riemann sphere are
allowed. The general solution to the conformal Killing equations is
$Y^\zeta=Y(\zeta)$, $Y^{\bar\zeta}=\bar Y(\bar\zeta)$, with $Y$ and
$\bar Y$ independent functions of their arguments. The standard basis
vectors are choosen as
\begin{equation}
l_n=-\zeta^{n+1}\frac{\d}{\d\zeta},\quad \bar l_n=-\bar
\zeta^{n+1}\frac{\d}{\d\bar \zeta},\quad n\in \mathbb Z\label{eq:55}
\end{equation}
At the same time, let us choose to expand the generators of the
supertranslations in terms of
\begin{equation}
  T_{m,n}=P^{-1}\zeta^m\bar\zeta^n, 
\quad m,n\in\mathbb Z. \label{eq:15}
\end{equation}
In terms of the basis vector $l_l\equiv (l_l,0)$ and
$T_{mn}=(0,T_{mn})$, the commutation relations for the complexified
$\mathfrak{bms}_4$ algebra read
\begin{equation}
\begin{gathered}
  \label{eq:37}
  [l_m,l_n]=(m-n)l_{m+n},\quad [\bar l_m,\bar l_n]=(m-n)\bar
  l_{m+n},\quad [l_m,\bar l_n]=0, \\
[l_l,T_{m,n}]=(\frac{l+1}{2}-m)T_{m+l,n},
\quad [\bar l_l,T_{m,n}]= (\frac{l+1}{2}-n)T_{m,n+l}. 
\end{gathered}
\end{equation}

The $\mathfrak{bms}_4$ algebra contains as subalgebra the Poincar\'e
algebra, which we identify with the algebra of exact Killing vectors
of the Minkowski metric equipped with the standard Lie bracket.

Indeed, these vectors form the subspace of spacetime vectors
\eqref{eq:26} for which (i) $\beta=0=U^A=\varphi$ while $V=-r$ and
$g_{AB}={}_0\gamma_{AB}$ and (ii) the relations in \eqref{eq:4b} hold
with $0$ on the right hand sides.  The former implies in particular
that $I^A=-\frac{1}{r}{}_0\gamma^{AB}\d_B f$, while a first
consequence of the latter is that the modified Lie bracket reduces the
standard one.

Besides the previous conditions that $Y^A$ is an $u$-independent
conformal Killing vector of the $2$-sphere,
$\cL_{Y}\,{}_0\gamma_{AB}={}_0D_C Y^C\,{}_0\gamma_{AB}$ and $f=T+\half
u\,{}_0\psi$ with $T_{,u}=0=T_{,r}$, we find the additional
constraints
\begin{gather}
  \label{eq:78}
  {}_0D_A \d_B\,{}_0\psi+{}_0D_B\d_A\,{}_0\psi={}_0
\gamma_{AB}\,{}_0\Delta\,{}_0\psi,\\
  {}_0D_A \d_B T+{}_0D_B\d_A T={}_0\gamma_{AB}\,{}_0\Delta T,\quad
  \d_A T=-\half \d_A({}_0\Delta T).
\end{gather}
In the coordinates $\zeta,\bar\zeta$, these constraints are equivalent
to $\d^3 Y=0=\bar\d^3 \bar Y$ and $\d^2\tilde T=0=\bar\d^2\tilde T$,
where $T=P^{-1}\tilde T$ and $\d=\frac{\d}{\d\zeta}$,
$\bar\d=\frac{\d}{\d\bar\zeta}$, so that the complexified Poincar\'e algebra is
spanned by the generators
\begin{equation}
  l_{-1},\,l_0,\,l_1,\quad \bar l_{-1},\, \bar l_0,\, \bar l_1,\quad
T_{0,0},\,T_{1,0},\,T_{0,1},\,T_{1,1},\label{eq:61}
\end{equation}
and the non vanishing commutation relations read
\begin{equation}
\begin{gathered}
  \label{eq:79}
[l_{-1},l_0]=-l_{-1},\ [l_{-1},l_1]=-2 l_0,\
[l_0,l_1]=-l_1,\\
[l_{-1},T_{1,0}]=-T_{0,0},\ [l_{-1},T_{1,1}]=-T_{0,1},\ [\bar
l_{-1},T_{0,1}]=-T_{0,0},\ [\bar l_{-1},T_{1,1}]=-T_{1,0},\\
[l_{0},T_{0,0}]=\half T_{0,0},\ [l_{0},T_{0,1}]=\half T_{0,1}, \
[l_{0},T_{1,0}]=-\half T_{1,0},\ [l_{0},T_{1,1}]=-\half T_{1,1},\\
[\bar l_{0},T_{0,0}]=\half T_{0,0},\ [\bar l_{0},T_{0,1}]=-\half T_{0,1}, \
[\bar l_{0},T_{1,0}]=\half T_{1,0},\ [\bar l_{0},T_{1,1}]=-\half T_{1,1},\\
[l_{1},T_{0,0}]= T_{1,0},\ [l_{1},T_{0,1}]= T_{1,1},\ [\bar
l_{1},T_{0,0}]=T_{0,1},\ [\bar l_{1},T_{1,0}]=T_{1,1}. 
\end{gathered}  
\end{equation} 
In particular for instance, the generators for translations can be
written as $\half(T_{1,1}+T_{0,0})=1$,
$\half(T_{1,1}-T_{0,0})=\cos\theta$, $\half ( T_{1,0}+
T_{0,1})=\sin\theta\cos\phi$, $\frac{1}{2i}( T_{1,0}-
T_{0,1})=\sin\theta\sin\phi$. Note that in order for the asymptotic
symmetry algebra to contain the Poincar\'e algebra as a subalgebra, it
is essential not to restrict the generators of supertranslations to
the sum of holomorphic and antiholomorphic functions on the Riemann
sphere.

The considerations above apply for all choices of $\varphi$ which is
freely at our disposal. In the original work of Bondi, van der Burg,
Metzner and Sachs, and in much of the subsequent work, the choice
$\varphi=0$ was preferred. From the conformal point of view, the
choice
\begin{equation}
\varphi=\ln{[\half(1+\zeta\bar\zeta)]}\label{eq:25}
\end{equation}
is interesting as it turns $\bar\gamma_{AB}$ into the flat metric on
the Riemann sphere with vanishing Christoffel symbols,
\begin{equation}
  \label{eq:9}
  \bar\gamma_{AB}dx^Adx^B=d\zeta d\bar\zeta. 
\end{equation}
In this case, $\psi =\d_A Y^A $,
\begin{equation}
f=\tilde T+\half u \psi,\label{eq:48}
\end{equation}
 with $\tilde
T=PT$. In terms of $\tilde T$, we get instead of the last of
\eqref{eq:32}
\begin{equation}
\hat{\tilde T}=Y^A_1\tilde T_2+\half \tilde T_1 \d_A
Y_2^A-(1\leftrightarrow 2).\label{eq:49}
\end{equation}
In terms of generators, the algebra \eqref{eq:37} is unchanged if one
now expands the supertranslations $\tilde T$ directly in terms of
$\tilde T_{m,n}=\zeta^m\bar\zeta^n$. 

More generally, one can also consider the transformations that leave
the form of the metric \eqref{eq:2} invariant up to a conformal
rescaling of $g_{AB}$, i.e., up to a rescaling of $\varphi$ by
$\omega(u,x^A)$. They are generated by spacetime vectors satisfying
\begin{gather}
  \label{eq:4aext}
  \cL_\xi g_{rr}=0,\quad \cL_\xi g_{rA}=0,\quad \cL_\xi g_{AB}
  g^{AB}=4\omega,\\
\cL_\xi
g_{ur}=O(r^{-2}),\quad \cL_\xi g_{uA}=O(1),\quad  \cL_\xi
g_{AB}=2\omega g_{AB}+O(r),\nonumber\\
\cL_\xi g_{uu}=-2r\dot\omega-2\omega e^{-2\varphi}+2\omega\bar\Delta
\varphi + O(r^{-1}). \label{eq:4bext}
\end{gather}

Equations \eqref{eq:4aext}, \eqref{eq:4bext} then imply that the
vectors are given by \eqref{eq:26} and \eqref{eq:44ter} with the
replacement $\psi\to \psi-2\omega$.

With this replacement, the vector fields $\bar\xi=f \dover{}{u}+
Y^A\dover{}{x^A}$ on $\Im={\mathbb R}\times S^2$ equipped with the
modified bracket provide a faithful representation of the extension of
$\mathfrak{bms}_4$ defined by elements $(Y,T,\omega)$ and bracket
$[(Y_1,T_1,\omega_1),(Y_2,T_2,\omega_2)]=(\hat Y,\hat T,\hat\omega)$,
with $\hat Y,\hat T$ as before and $\hat \omega=0$.

Indeed, the result is obvious for the $A$ components. Furthermore,
\[\delta^g_{\bar \xi_1} f_2= \omega_1 f_2+\half e^{\varphi}\int_0^udu^\prime
e^{-\varphi}[-\omega_1(\psi_2-2\omega_2)+
2Y^A_2\d_A \omega_1].\] At $u=0$, we get
$[\bar\xi_1,\bar\xi_2]^u_M|_{u=0}=e^\varphi|_{u=0} \hat T$, while
direct computation shows that $\d_u ([\bar\xi_1,\bar\xi_2]^u_M)=\hat
f\dot\varphi +\half \bar D_B\hat Y^B$, as it should. 

Following the same reasoning as before, one can then also show that
the spacetime vectors \eqref{eq:26} with the replacement
discussed above and equipped with the modified Lie bracket provide a
faithful representation of the extended $\mathfrak{bms}_4$ algebra. 

Indeed, we have $\xi=\bar\xi +I^A\dover{}{x^A}+\xi^r\dover{}{r}$.
Furthermore, $[\xi_1,\xi_2]_M^u=[\bar\xi_1,\bar\xi_2]^u_M=\hat f$ as
it should.  In the extended case, the variations of $\beta,U^A$ are
still given by \eqref{eq:10bis}.  We then have
$\lim_{r\to\infty}[\xi_1,\xi_2]_M^A=\hat Y^A$ and find, after some
computations, $\d_r ([\xi_1,\xi_2]_M^A)=g^{AB}e^{2\beta} \d_B\hat f$,
giving the result for the $A$ components.  Finally, for the $r$
component, we find
$\lim_{r\to\infty}\frac{[\xi_1,\xi_2]^r_M}{r}=-\half (\hat \psi+2\hat
f \dot\varphi)$, while $\d_r(\frac{[\xi_1,\xi_2]^r_M}{r})=-\half (\d_r
\hat \chi-\d_B\hat f\d_r U^B)$, which concludes the proof.

In order to make contact with the original literature, we have choosen
the conformal factor with respect to the unit sphere, $\bar
\gamma_{AB}=e^{2\varphi}\,{}_0\gamma_{AB}$. Computations could have
been simplified and the derivation of algebra \eqref{eq:37} would have
been streamlined by introducing the conformal factor directly with
respect to the flat metric,
$\bar\gamma_{AB}dx^Adx^B=e^{2\tilde\varphi}d\zeta d\bar\zeta$, with
$\tilde \varphi(u,\zeta,\bar\zeta)=\varphi-\ln P$, as presented in
\cite{Barnich:2009se}. In this case, the determinant condition is taken
as $b=\frac{1}{4}e^{4\tilde\varphi}$, while the boundary condition
that involves the conformal factor now reads
\begin{equation}
\frac{V}{r}=-2r\d_u \tilde\varphi+\bar\Delta
  \tilde\varphi+O(r^{-1})\label{eq:7},
\end{equation}
where $\bar \Delta \tilde \varphi=4
e^{-2\tilde\varphi}\d\bar\d\tilde\varphi$ with $\d=
\d_\zeta,\bar\d=\d_{\bar\zeta}$. 

\subsection{Solution space}

We start by assuming only that we have a metric of the form
\eqref{eq:2} and that the determinant condition holds. 
Following again \cite{Tamburino:1966}, the equations of motion are
organized in terms of the Einstein tensor
$G_{\alpha\beta}=R_{\alpha\beta}-\half g_{\alpha\beta} R$ as
\begin{gather}
  G_{r\alpha}=0,  \qquad
G_{AB}-\half g_{AB} g^{CD}G_{CD}=0,  \label{eq:57b}\\
G_{uu}=0=G_{uA},  \label{eq:57c}\\
g^{CD}G_{CD}=0.  \label{eq:57d} 
\end{gather}
Due to the form of the metric and the determinant condition, equation
\eqref{eq:57d} is a consequence of \eqref{eq:57b} on
account of the Bianchi identities. Indeed, the latter can be written
as
\begin{equation}
\label{eq:56}
 0= 2\sqrt{-g}{G_\alpha^\beta}_{;\beta}=2(\sqrt{-g}G_\alpha^\beta)_{,\beta}+
\sqrt{-g}G_{\beta\gamma}{g^{\beta\gamma}}_{,\alpha}. 
\end{equation}
When \eqref{eq:57b} hold and $\alpha=1$, we get
$G_{AB}{g^{AB}}_{,r}=0=\half g_{AB}{g^{AB}}_{,r} g^{CD}G_{CD}$. This
implies that \eqref{eq:57d} holds by using \eqref{eq:27}. 

The remaining Bianchi identities then reduce to
$2(\sqrt{-g}G_A^\beta)_{,\beta}=0=2(\sqrt{-g}G_u^\beta)_{,\beta}$. The
first gives $(r^2 G_{uA})_{,r}=0$. This means that if $r^2G_{uA}=0$
for some constant $r$, it vanishes for all $r$. When $G_{uA}=0$, the
last Bianchi identity reduces to $(r^2G_{uu})_{,r}=0$, so that again,
$r^2G_{uu}=0$ everywhere if it vanishes for some fixed $r$.

Let $k_{AB}=\half g_{AB,r}$, $l_{AB}=\half g_{AB,u}$, $n_A=
\frac{1}{2} e^{-2\beta}g_{AB} U^B_{,r}$ with indices on these
variables and on $U^A$ lowered and raised with the 2 dimensional
metric $g_{AB}$ and its inverse.  Define $K^A_B$ through the relation
$k^A_B=\frac{1}{r}\delta^A_B+\frac{1}{r^2}K^A_B$.  In particular, the
determinant condition implies that $k=\frac{2}{r}$ and thus that
$K^D_D=0$.  Similarily, if
$l^D_B=\half\bar\gamma^{DA}\bar\gamma_{AB,u}+\frac{1}{r}L^D_B$, the
determinant condition implies in particular that $L^D_B$ is traceless,
$L^D_D=0$. Note that for a traceless $2\times 2$ matrix ${M^T}^A_B$,
we have ${M^T}^A_C{M^T}^C_B=\half {M^T}^C_D{M^T}^D_C \delta^A_B$.

For a metric of the form \eqref{eq:2}, we have 
\begin{equation*}
\begin{gathered}
\Gamma^\lambda_{rr}=\delta^\lambda_r2\beta_{,r},\quad
\Gamma^{u}_{\lambda r}=0,\quad
\Gamma^r_{Ar}=\beta_{,A}+ n_A,\quad \Gamma^A_{Br}=k^A_B,\\
\Gamma^u_{AB}= e^{-2\beta} k_{AB},\quad
\Gamma^A_{BC}= e^{-2\beta} U^A k_{BC}+{}^{(2)}\Gamma^A_{BC}, \\ \Gamma^A_{ur}=
-k^A_BU^B+e^{2\beta}(\d^A\beta-n^A),\quad 
\Gamma^u_{uA}=\beta_{,A}-n_A- e^{-2\beta} k_{AB}U^B,\\
\Gamma^r_{ur}=-\half (\d_r+2\beta_{,r})\frac{V}{r}-(\beta_{,A}+n_A) U^A,\\
\Gamma^A_{Bu}=l^A_B +\half {}^{(2)}D^A U_B-\half {}^{(2)}D_B U^A+U^A(\beta_{,B}-n_B)- e^{-2\beta}k_{BC}U^A
U^C,\\
\Gamma^{u}_{uu}=2\beta_{,u}+\half(\d_r+2\beta_{,r})
\frac{V}{r}+2U^A n_A+e^{-2\beta}k_{AB} U^AU^B,\\
\Gamma^r_{AB}= e^{-2\beta}(\half{}^{(2)}D_A U_B
+\half{}^{(2)}D_B U_A+l_{AB}+k_{AB}\frac{V}{r} ),\\
\Gamma^r_{uA}=-\big(\frac{V_{,A}}{2r}+\frac{V}{r}n_A+
e^{-2\beta}U^B[\frac{1}{2}{}^{(2)}D_A U_B+\frac{1}{2}{}^{(2)}D_B
U_A+l_{AB}+\frac{V}{r}k_{AB}]\big),\\
\Gamma^A_{uu}=2U^A\beta_{,u}+\half
U^A(\d_r+2\beta_{,r})\frac{V}{r}+2U^A n_B U^B+U^A k_{BC}e^{-2\beta}U^BU^C
\\-U^A_{,u}-2l^A_B U^B-\half
e^{2\beta}(\d^A+2\d^A\beta)\frac{V}{r}-\half {}^{(2)}D^A(U^CU_C),\\
\Gamma^r_{uu}=-\half(\d_u-2\beta_{,u})\frac{V}{r}+\half
\frac{V}{r}(\d_r+2\beta_{,r})\frac{V}{r}+\half
U^A(\d_A+2\beta_{,A})\frac{V}{r}+ 2\frac{V}{r} U^A n_A\\+\frac{V}{r}
e^{-2\beta}k_{AB} U^AU^B+e^{-2\beta}l_{AB}U^AU^B +e^{-2\beta} U^AU^B
{}^{(2)}D_A U_B. 
\end{gathered}
\end{equation*}

To write the equations of motion, we use that $|{}^{(4)} g|=e^{4\beta}
|{}^{(2)}g|$ and
\begin{equation*}
  R_{\mu\nu}=\big[\d_\alpha+(2\beta+\half\ln
  |{}^{(2)}g|)_{,\alpha}\big]\Gamma^\alpha_{\mu\nu}-\d_\mu\d_\nu(2\beta+\half\ln
  |{}^{(2)}g|) -\Gamma^\alpha_{\nu\beta}\Gamma^\beta_{\mu\alpha}. 
\end{equation*}
The equation $G_{rr}\equiv R_{rr}=0$ then becomes
\begin{equation}
  \label{eq:60}
  \d_r\beta=-\frac{1}{2r}+\frac{r}{4} k^A_B k^B_A=\frac{1}{4r^3} K^A_B K^B_A\iff
  \beta=-\int^\infty_r dr^\prime\frac{1}{4 {r^\prime}^3} K^A_B K^B_A\,.
\end{equation}
This equation determines $\beta$ uniquely in terms of $g_{AB}$ because
the fall-off condition \eqref{eq:3b} excludes the arbitrary function of
$u,x^A$ allowed by the general solution to this equation. 

The equations $G_{rA}\equiv R_{rA}=0$ read
\begin{equation}
\begin{gathered}
  \label{eq:54}
  \d_r(r^2n_A)=J_A, \\
  J_A=r^2(\d_r -\frac{2}{r})\beta_{,A}-{}^{(2)}D_B K^B_A
  =\d_A(-2r\beta+\frac{1}{4r}K^B_CK^C_B)-{}^{(2)}D_B K^B_A. 
\end{gathered}
\end{equation}
In the original approach \cite{Bondi:1962px,Sachs:1962wk}, it was
assumed in particular that the metric $g_{AB}$ admits an expansion in
terms of powers of $r^{-1}$ starting at order $r^2$. We will assume
\begin{equation}
  \label{eq:96}
  g_{AB}=r^2\bar\gamma_{AB}+rC_{AB}+D_{AB}+\frac{1}{4} \bar\gamma_{AB}
C^C_DC^D_C+o(r^{-\epsilon}), 
\end{equation}
where indices on $C_{AB},D_{AB}$ are raised with the inverse of
$\bar\gamma_{AB}$.  In \cite{Tamburino:1966}, it was then shown
explicitly how \eqref{eq:96} is related to the conformal approach
\cite{PhysRevLett.10.66,Penrose:1965am} and imposed through
differentiability conditions at null infinity.

Under the assumption \eqref{eq:96}, $C^D_D=0=D^C_C$ and
\begin{equation}
\begin{split}
  \label{eq:97}
  K^A_B&=-\half C^A_B- r^{-1}D^A_B+o(r^{-1-\epsilon}),\\
  \beta&=-\frac{1}{32}r^{-2}C^A_BC^B_A-\frac{1}{12}
    r^{-3} C^A_BD^B_A+o(r^{-3-\epsilon}),\\
  J_A&=\half \bar D_BC^B_A+ r^{-1} \bar
  D_BD^B_A +o(r^{-1-\epsilon}).
\end{split}
\end{equation}
These equations then imply
$n_A=\frac{1}{2}r^{-1}\bar D_BC^B_A+r^{-2} (\ln r \bar
D_BD^B_A+N_A)+ o(r^{-2-\epsilon})$ and involve the
arbitrary functions $N_A(u,x^B)$ as integration ``constants''.
Because $U^A$ has to vanish for $r\to\infty$, we get from the
definition of $n_A$
\begin{equation}
  \label{eq:62}
  U^A=-\frac{1}{2}r^{-2}\bar D_BC^{BA}-
\frac{2}{3}r^{-3}\Big[(\ln r+
  \frac{1}{3})\bar
D_BD^{BA}-\half C^{A}_{B} \bar D_CC^{CB}
  +N^A\Big]
  +o(r^{-3-\varepsilon}),
\end{equation}
where the index on $N_A$ has been raised with $\bar\gamma^{AB}$. 

It is straightforward to verify that if one trades the coordinate $r$
for $s=r^{-1}$, the only non vanishing components of the
``unphysical'' Weyl tensor at the boundary are given by 
\begin{equation}
\lim_{s\to
  0}(s^2W_{sAsB})=-D_{AB},\label{eq:4}
\end{equation}
(see e.g.~\cite{Winicour:1985pi} for a
detailed discussion). In \cite{Sachs:1962wk}, the condition $D_{AB}=0$
was imposed in order to avoid a logarithmic $r$-dependence in the
solution to the equations of motion and to avoid singularities on the
unit sphere. When one dispenses with this latter restriction, absence
of a logarithmic $r$-dependence is guaranteed through the requirement
$\bar D_BD^{BA}=0$. In the coordinates $\zeta,\bar\zeta$ and with the
parametrization $\bar\gamma_{AB}dx^Adx^B=e^{2\tilde\varphi}d\zeta
d\bar\zeta$, this is equivalent to
\begin{equation}
  \label{eq:98}
  D_{\zeta\zeta}=d(u,\zeta), \quad D_{\bar
  \zeta\bar \zeta}=\bar d(u,\bar \zeta), \quad 
D_{\zeta\bar \zeta}=0. 
\end{equation}
A more complete analysis of the field equations when allowing for a
logarithmic or, more precisely, a ``polyhomogeneous'' dependence in $r$
can be found in \cite{Chrusciel:1993hx}.

Starting from 
\begin{multline*}
  R_{AB}= (\d_r+2\beta_{,r}+\frac{2}{r})\Gamma^r_{AB}-
k^C_A\Gamma^r_{BC}-k^C_B\Gamma^r_{AC}
+{}^{(2)}R_{AB}-2{}^{(2)}D_B\beta_{,A}
 \\+(\d_u+2\beta_{,u}+l )\Gamma^u_{AB}-\Gamma^u_{uA}\Gamma^u_{uB}
-\Gamma^r_{rA}
  \Gamma^r_{rB}\\
-\Gamma^C_{uA}\Gamma^u_{BC}
  -\Gamma^C_{uB}\Gamma^u_{AC}+ {}^{(2)}D_C( e^{-2\beta}U^C
  k_{AB})\\-e^{-4\beta}U^C k_{BD}U^D
  k_{AC}+2
   e^{-2\beta}\beta_{,C}U^C k_{AB},
\end{multline*}
we find 
\begin{multline}
  \label{eq:72}
  g^{DA}R_{AB}=e^{-2\beta}
\Big[(\d_r+\frac{2}{r})(l^D_B+k^D_B\frac{V}{r}+\half{}^{(2)}D_B
  U^D+\half {}^{(2)}D^D U_B)\\+k^D_A{}^{(2)}D_B
  U^A-k^A_B{}^{(2)}D_A U^D+(\d_u+l)k^D_B+{}^{(2)}D_C (U^C k^D_B)\Big]\\+{}^{(2)}R^D_{B}
  -2({}^{(2)}D_B\d^D\beta+\d^D\beta\d_B\beta +n^Dn_B). 
\end{multline}

When taking into account the previous equations, $G_{ur}\equiv
R_{ur}+\half e^{2\beta} R=0$ reduces to $g^{AB}R_{AB}=0$. Explicitly,
we find from the trace of \eqref{eq:72}
\begin{equation}
\begin{gathered}
  \label{eq:68}
  \d_r V=J,\\
  J=e^{2\beta}r^2({}^{(2)}\Delta
  \beta+\d^D\beta\d_D\beta+n^D n_D-\half{}^{(2)}R)
-2rl -\frac{r^2}{2}(\d_r+\frac{4}{r}) {}^{(2)}
  D_B U^B\\
  =-2rl+
e^{2\beta}r^2\big[{}^{(2)}\Delta
  \beta+(n^A-\d^A\beta)(n_A-\d_A\beta)\\-{}^{(2)}D_A
  n^A-\half{}^{(2)}R\big] -2r {}^{(2)} D_B U^B
\\=-2rl-\frac{1}{2}\bar R+o(r^{-1-\epsilon}), 
\end{gathered}
\end{equation}
where we have used the previous equation to get the second line. 
This equation implies
\begin{equation}
  \label{eq:69}
  \frac{V}{r}  =-r l-\frac{1}{2} \bar
  R + r^{-1}2M + o(r^{-1-\epsilon}),
\end{equation}
and implies a third arbitrary function of $M(u,x^B)$ as integration
constant.

We have $G_{AB}-\half g_{AB} g^{CD} G_{CD}=R_{AB}-\half g_{AB} g^{CD}
R_{CD}$. Taking into account the previous equations, it thus reduces
to the condition that the traceless part of \eqref{eq:72} vanishes.
Using that $\d_u k^D_B=\d_r l^D_B-2(l^D_Ak^A_B-k^D_A
l^A_B)$, we get 
\begin{multline*}
  (\d_r+\frac{1}{r}) l^D_B-
  (l^D_Ak^A_B-k^D_Al^A_B)+\frac{1}{2}k^D_B l =\\
-\half \Big[ (\d_r+\frac{2}{r})(k^D_B\frac{V}{r}+\half{}^{(2)}D_B
  U^D+\half {}^{(2)}D^D U_B)\\+k^D_C{}^{(2)}D_B
  U^C-k^C_B{}^{(2)}D_C U^D+{}^{(2)}D_C (U^C k^D_B)\Big]\\+e^{2\beta} \Big[ n^Dn_B+
  {}^{(2)}D_B\d^D\beta+\d^D\beta\d_B\beta -\half {}^{(2)}R^D_{B}\Big]. 
\end{multline*}
The various definitions then give 
\begin{equation}
  \label{eq:74}
  \d_r L^D_B-\frac{1}{r^2}
  (L^D_AK^A_B-K^D_AL^A_B)=J^D_B,
\end{equation}
where
\begin{multline}
J^D_B=-\frac{r}{2} \Big[ (\d_r+\frac{2}{r})(k^D_B\frac{V}{r}+\half{}^{(2)}D_B
  U^D+\half {}^{(2)}D^D U_B)\\+k^D_C{}^{(2)}D_B
  U^C-k^C_B{}^{(2)}D_C U^D+{}^{(2)}D_C (U^C k^D_B)\Big]+\\+r e^{2\beta} \Big[ n^Dn_B+
  {}^{(2)}D_B\d^D\beta+\d^D\beta\d_B\beta-\half {}^{(2)}R^D_{B}\Big]-\frac{1}{2}
  \bar\gamma^{DA}
  \bar\gamma_{AB,u}-\frac{r}{2}k^D_B l
  \\-\frac{1}{2r}
  (K^D_A\bar\gamma^{AC}\bar\gamma_{CB,u}-
\bar\gamma^{DC}\bar\gamma_{CA,u}
  K^A_B).
\end{multline}
The previous equations imply
\begin{multline*}
  J^D_B=-\half (\d_r k^D_B+\frac{1}{r}k^D_B)V-\frac{r^2}{2}k^D_Be^{2\beta}\big[{}^{(2)}\Delta
  \beta+(n^A-\d^A\beta)(n_A-\d_A\beta)-{}^{(2)}D_A n^A-\half{}^{(2)}R\big]\\
-\half({}^{(2)}D_B
  U^D+ {}^{(2)}D^D U_B)-r  U^C {}^{(2)}D_C k^D_B+\frac{r}{2} k^D_C({}^{(2)}D^C
  U_B- {}^{(2)}D_B U^C)\\+\frac{r}{2}k^C_B({}^{(2)}D_C U^D-{}^{(2)}D^D
  U_C)+\frac{r}{2} {}^{(2)}D_C U^Ck^D_B
\\+r e^{2\beta} \Big[ (n^D-\d^D\beta)(n_B-\d_B\beta)+
  {}^{(2)}D_B\d^D\beta-\half ({}^{(2)}R^D_{B}+{}^{(2)}D_B
  n^D+ {}^{(2)}D^D n_B)\Big]\\-\frac{1}{2}
  \bar\gamma^{DA}
  \bar\gamma_{AB,u}+\frac{r}{2}k^D_Bl
 -\frac{1}{2r}
  (K^D_A\bar\gamma^{AC}\bar\gamma_{CB,u}-
\bar\gamma^{DC}\bar\gamma_{CA,u}
  K^A_B).
\end{multline*}

Let $\cO^{DA}_{BC}=-\frac{1}{r^2} (K^D_C\delta^A_B-\delta^D_CK
^A_B)$ and
$\cA\cR$ denote anti-radial ordering. 
Equation \eqref{eq:74} without right-hand side has the same form as
the Schr\"odinger equation with time dependent Hamiltonian. If we define
\begin{equation}
U^{DA}_{BC}(r_<,r_>)=\cA\cR\exp{[-\int^{r_>}_{r_{<}}
  dr^\prime \cO^{DA}_{BC}(r^\prime)]}\label{eq:76},
\end{equation}
the solution to the inhomogeneous equation \eqref{eq:74} with
non-vanishing $J^B_D$ can then be obtained by variation of constants
and reads
\begin{equation}
  \label{eq:75}
  L^D_B(r)=U^{DA}_{BC}(r,\infty)[\half N^C_A+\int dr^\prime
  U^{CE}_{AF}(\infty,r^\prime) J^F_E(r^\prime)], 
\end{equation}
and involves two more integration constants encoded in $N^D_B(u,x^B)$.

In other words, the $r$-dependence of $g_{AB,u}$ is completely
determined up to two integration constants. It follows that the only
variables left in the theory whose $r$-dependence is undetermined are
the two functions contained in $E_{AB}(u_0,r,x^C)=g_{AB}(u_0,r,x^C)
-r^2\bar\gamma_{AB}(u_0,x^C)-rC_{AB}(u_0,x^C)
-D_{AB}(u_0,x^C)-\frac{1}{4}\bar\gamma_{AB} C^C_DC^D_C$ at some
  initial fixed $u_0$.

When expanding into orders in $r$, one finds in particular
\begin{equation*}
\begin{split}
  L^D_B&=\half(\bar\gamma^{DA}C_{AB,u}-
  C^{DA}\bar\gamma_{AB,u} )+\half
  r^{-1}\Big[\bar\gamma^{DA}\d_u (D_{AB}+\frac{1}{4}\bar\gamma_{AB}C^C_DC^D_C)
\\&  -C^{DA}C_{AB,u}
  -D^{DA}\bar\gamma_{AB,u}
  +\frac{1}{4}C^E_FC^F_E
  \bar\gamma^{DA}\bar\gamma_{AB,u}\Big]+o(r^{-1-\epsilon}),
  \\
  J^D_B&= \half \delta^D_Bl-\half\bar\gamma^{DA}\bar\gamma_{AB,u} +
  \frac{1}{4}r^{-1}[C^{DA}\bar\gamma_{AB,u}
  -\bar\gamma^{DC}\bar\gamma_{CA,u} C^A_B]\\&+\half
  r^{-2}[ lD^D_B+
  D^{DA}\bar\gamma_{AB,u}-\bar\gamma^{DC}
  \bar\gamma_{CA,u}D^A_B] +o(r^{-2+\epsilon}).
\end{split}
\end{equation*}
When injecting into the equation of motion \eqref{eq:74}, the leading
order requires that 
\begin{equation}
  \label{eq:101}
  \bar\gamma_{AB,u}=l\bar\gamma_{AB}, 
\end{equation}
or, in other words, that the only $u$ dependence in $\bar\gamma_{AB}$
is contained in the conformal factor. This agrees with the assumption
of section~\bref{sec:asympt-flat-4}, where the $u$-dependence of
$\bar\gamma_{AB}$ was contained in $\exp{2\varphi}$ and 
$l=2\d_u\varphi$, and also with the
discussion at the end of the previous subsection, where it was
contained in $\exp{2\tilde\varphi}$ and $l=2\d_u\tilde\varphi$. In
the following we always assume that 
\eqref{eq:101} holds. In particular, this implies
\begin{equation*}
\begin{split}
  L^D_B&=\half(\bar\gamma^{DA}C_{AB,u}- lC^D_B
  )+\half r^{-1}[\bar\gamma^{DA}D_{AB,u}
  -C^{DA}C_{AB,u} \\ &-lD^D_B
  +\frac{1}{2}C^{EF}\d_uC_{EF}
  \delta^D_B]+o(r^{-1-\epsilon}),
  \\
  J^D_B&= \half r^{-2} lD^D_B+o(r^{-2+\epsilon}).
\end{split}
\end{equation*}
When taking into account the next order of \eqref{eq:74} and comparing
to the general solution \eqref{eq:75}, we get 
\begin{equation}
  \label{eq:99}
  \d_u D_{AB}=0, \qquad 
N_{AB}=\d_u C_{AB}-C_{AB} l,
\end{equation}
where the index on $N^A_B$ has been lowered with
$\bar\gamma_{AC}$. This implies in turn that 
\begin{equation*}
  l^A_B=\half l\delta^A_B+\half r^{-1} N^A_B-\frac{1}{4} r^{-2}[
 C^{A}_{C}N^{C}_{B}- N^A_CC^C_B
+2lD^A_B] +o(r^{-2-\epsilon}). 
\end{equation*}

At this stage, equations \eqref{eq:57b} have been solved, and then
\eqref{eq:57d} holds automatically on account of the Bianchi
identities. Furthermore $g^{CD}G_{CD}=0$ reduces to $R_{ur}=0$ and we
also have $R=0$.  Under these assumptions, we only need to discuss the
$r$ independent part of $r^2G_{uA}=0$ and then of $r^2G_{uu}=0$, which
reduce to $r^2 R_{uA}=0$ and $r^2 R_{uu}=0$, respectively. The
$r$-independent part fixes the $u$ dependence of $N_A$ and $M$ in
terms of the other fields. Explicitly,
\begin{multline*}
  R_{uA}=(-\d_u+l)\beta_{,A} -\d_A l-(\d_u+l)n_A  
 \\ +n_B {}^{(2)}D^B U_A- \beta_{,B}  {}^{(2)}D_A
  U^B  +2U^B(\beta_{,B}\beta_{,A}+n_Bn_A)
\\+ {}^{(2)}D_B
  \Big[l^B_A+\half  {}^{(2)}D^B U_A
-\half  {}^{(2)}D_A
  U^B+U^B(\beta_{,A}-n_A)\Big]+2n_Bl^B_A
\\  -(\d_r+2\beta_{,r}+\frac{2}{r}
  )(\frac{V_{,A}}{2r})-\frac{V}{r}(\d_r+\frac{2}{r}
  )  n_A+k_A^B(\frac{V_{,B}}{r}+2\frac{V}{r} n_B )
\\-e^{-2\beta} (\d_r+\frac{2}{r})\big[U^B(\half
   {}^{(2)}D_A U_B+\half  {}^{(2)}D_B U_A+l_{AB}+\frac{V}{r}k_{AB})\big]
\\
-e^{-2\beta}U^B\Big[(\d_u+l)k_{AB}-2l_A^Ck_{CB}-2k_A^Cl_{CB}
-2k_A^Ck_{CB}\frac{V}{r}\\+ {}^{(2)}D_C(k_{AB}
  U^C)-k_{AC} {}^{(2)}D^C U_B-k_{BC} {}^{(2)}D^C U_A\Big],
\end{multline*}
and the term proportional to $r^{-2}$ yields
\begin{multline}
  \label{eq:78ter}
(\d_u+l)N_A
 =\d_AM+\frac{1}{4}C_A^B\d_B\bar
  R +\frac{1}{16}\d_A\big[N^B_C C^C_B\big]
-\frac{1}{4} \bar D_AC^C_BN^B_C\\
-\frac{1}{4}\bar D_B\big[C^{B}_{C}N^C_{A}-N^B_CC^C_A\big]-\frac{1}{4}
  \bar D_B \big[ \bar D^B \bar D_CC^C_A -\bar D_A \bar
  D_CC^{BC}\big]\\-\frac{1}{32}
l\d_A\big[C^B_CC^C_B\big]
 +\frac{1}{16}\d_A lC^B_CC^C_B 
+\half \bar D_B\big[lD^B_A\big].
\end{multline}
Similarily, 
\begin{multline*}
   R_{uu}=(\d_u+2\beta_{,u}+l)\Gamma^u_{uu}+(\d_r+2\beta_{,r}+
\frac{2}{r})\Gamma^r_{uu} +
(\d_A+2\beta_{,A}+{}^{(2)}\Gamma^{B}_{BA})\Gamma^A_{uu} \\-2\beta_{,uu}-\d_u l 
-(\Gamma^{u}_{uu})^2-2\Gamma^{u}_{uA}\Gamma^A_{uu}-(\Gamma^r_{ur})^2 
-2\Gamma^r_{uA}\Gamma^A_{ur}-\Gamma^A_{uB}\Gamma^B_{uA},
\end{multline*}
and the term proportional to $r^{-2}$ yields
\begin{multline}
  \label{eq:3}
  (\d_u+\frac{3}{2}l) M=-\frac{1}{8} N^A_BN^B_A
  -\frac{1}{8}lC^A_BN^B_A-\frac{1}{32} l^2 C^A_BC^B_A +\frac{1}{8}
  \bar \Delta \bar R \\
  +\frac{1}{4}\bar D_A\bar D_C N^{CA} +\frac{1}{8} l \bar D_A \bar D_C
  C^{CA}+\frac{1}{4}\bar D_C l\bar D_A C^{CA}.
\end{multline}
All these considerations can be summarized as follows:

{\em For a metric of the form \eqref{eq:2} satisfying the determinant
  condition and with $g_{AB}$ as in \eqref{eq:96}, the general
  solution to Einstein's equations is parametrized by the $2$
  dimensional background metric $\bar\gamma_{AB}(u,x^C)$ satisfying
  \eqref{eq:101}, by the mass and angular momentum aspects
  $M(u,x^A),N_A(u,x^B)$ satisfying \eqref{eq:3},\eqref{eq:78ter}, by
  the traceless symmetric news tensor $N_{AB}(u,x^C)$ defined in
  \eqref{eq:99}, and by the traceless symmetric tensors $D_{AB}(x^C)$,
  $C_{AB}(u_0,x^C)$, $E_{AB}(u_0,r,x^C)$.

  For such spacetimes, the only non vanishing components of the
  unphysical Weyl tensor at the boundary are given by \eqref{eq:4}.
  When logarithmic terms are required to be absent in the metric,
  $D_{AB}(x^C)$ has to satisfy $\bar D_BD^B_A=0$. In the coordinates
  $\zeta,\bar\zeta$ and the parametrization
  $\bar\gamma_{AB}dx^Adx^B=e^{2\tilde\varphi} d\zeta d\bar \zeta$,
  this leads to \eqref{eq:98} with $d=d(\zeta)$ and $\bar d=\bar d(\bar
\zeta)$ by also taking \eqref{eq:99} into account.}

In particular, let us now use the parametrization
$\bar\gamma_{AB}dx^Adx^B=e^{2\tilde\varphi}d\zeta d\bar\zeta$. The
determinant condition then reads $\text{det}\, g_{AB}=e^{4\tilde
  \varphi}\frac{r^4}{4}$.  Even though we will not use it explicitly
below, let us point out that the determinant condition can be
implemented for instance by choosing the Beltrami representation,
\begin{equation*}
\begin{gathered}
    h=\frac{g_{\zeta\zeta}}{g_{\zeta\bar\zeta}+f},\quad 
  \bar h=\frac{g_{\bar\zeta\bar\zeta}}{g_{\zeta\bar\zeta}+f}, \\
  g_{\zeta\zeta}=\frac{2f h}{1-y},\quad g_{\bar\zeta\bar\zeta}=
\frac{2f \bar h}{1-y}, \quad g_{\zeta\bar\zeta}=\frac{f(1+y)}{1-y},\\
  g^{\zeta\zeta}=-\frac{2 \bar h}{f(1-y)},\quad
  g^{\bar\zeta\bar\zeta}=-\frac{2 h}{f(1-y)},\quad
  g^{\zeta\bar\zeta}=\frac{1+y}{f(1-y)},
\end{gathered}
\end{equation*}
where $f=\sqrt{-{}^{(2)}g}$, $y=h\bar h$, with
$f=\frac{r^2}{2}e^{2\tilde\varphi}$ fixed, while $h=O(r^{-1})=\bar
h$. Alternatively, one can choose
\begin{equation*}
\begin{gathered}
   g_{\zeta\zeta}=f e^{i\alpha}\sinh\rho ,\quad g_{\bar\zeta\bar\zeta}=
f e^{-i\alpha}\sinh\rho, \quad g_{\zeta\bar\zeta}=f \cosh\rho,\\
  g^{\zeta\zeta}=-f^{-1}e^{-i\alpha}\sinh\rho, \quad
  g^{\bar\zeta\bar\zeta}=-f^{-1}e^{i\alpha}\sinh\rho ,\quad
  g^{\zeta\bar\zeta}=f^{-1}\cosh\rho,
\end{gathered}
\end{equation*}
where $\rho=O(r^{-1})$ and $\alpha=O(r^0)$.  

In the parametrization with the conformal factor introduced with
respect to the Riemann sphere, we can write
\begin{equation}
  \label{eq:17}
\begin{gathered}
  C_{\zeta\zeta}=e^{2\tilde\varphi} c,\quad
  C_{\bar\zeta\bar\zeta}=e^{2\tilde\varphi} \bar c,\quad
  C_{\zeta\bar\zeta}=0,\\
D_{\zeta\zeta}= d,\quad
  D_{\bar\zeta\bar\zeta}= \bar d,\quad
  D_{\zeta\bar\zeta}=0.
\end{gathered}
\end{equation}
Equations \eqref{eq:97}, \eqref{eq:62} and \eqref{eq:69} read 
\begin{equation}
\begin{split}
  \label{eq:30}
  \beta&=-\frac{1}{4}r^{-2}c\bar c -\frac{1}{3}r^{-3}e^{-2\tilde\varphi}
  (d\bar  c+\bar d c) +o(r^{-3-\epsilon}), \\
  U^\zeta&=-\frac{2}{r^{2}}e^{-4\tilde\varphi}\d(e^{2\tilde\varphi}\bar
  c)-\\&\hspace{2cm}-\frac{2}{3r^3}\Big[(\ln r+\frac{1}{3})4e^{-4\tilde\varphi}\d
  \bar d -4e^{-4\tilde\varphi}\bar c
\bar \d(e^{2\tilde\varphi} c)+ N^\zeta\Big]+o(r^{-3-\epsilon}),\\
  \frac{V}{r}&=-2r\d_u\tilde\varphi + 4e^{-2\tilde\varphi}\d\bar\d
  \tilde\varphi + r^{-1}2M +o(r^{-1-\epsilon}).
\end{split}
\end{equation}
and the evolution equations become
\begin{equation}
\begin{gathered}
  \label{eq:31}
  \d_u (e^{3\tilde\varphi} M)=\d_u \Big(e^{\tilde\varphi}\big[\d^2\bar
  c+\bar \d^2 c+2\d\tilde\varphi \d\bar c+2\bar\d\tilde\varphi\bar \d
  c+2\d^2\tilde\varphi\bar c+2\bar\d^2\tilde\varphi
  c\big]\Big)\\
  -e^{\tilde\varphi}\d_u(e^{\tilde\varphi}c)\d_u(e^{\tilde\varphi}\bar
  c)+2e^{\tilde\varphi}\Big( c\big[\d_u(\bar\d \tilde\varphi)^2
  -\d_u\bar\d^2\tilde\varphi\big] +\bar
  c\big[\d_u(\d\tilde\varphi)^2-\d_u\d^2\tilde\varphi\big]\Big)
  \\+e^{-\tilde\varphi}\Big(-4(\d\bar\d)^2\tilde\varphi+8\big[
  (\d\bar\d \tilde\varphi)^2+\d\tilde\varphi\d\bar\d^2\tilde\varphi+
  \bar\d\tilde\varphi\d^2\bar\d\tilde\varphi -2\bar\d\tilde\varphi
  \d\tilde\varphi\d\bar\d\tilde\varphi\big]\Big),
  \\
  \d_u (e^{2\tilde\varphi} N_{\bar\zeta})=e^{2\tilde\varphi}\Big[\bar
  \d M +\frac{1}{4}\big[(\bar\d \bar c+5\bar c\bar\d)\d_u c-(3
  c\bar\d+7\bar\d c) \d_u\bar c \big] +2\bar\d\tilde\varphi(\bar c\d_u
  c - c\d_u\bar c) \\-\half\d_u\tilde\varphi \bar\d (c\bar
  c)+\bar\d\d_u\tilde\varphi c\bar c \Big]
  +2(\d\d_u\tilde\varphi +\d_u\tilde\varphi\d)\bar d \\
  +\bar \d^3c+2\bar \d^3\tilde\varphi c +4\bar \d^2\tilde\varphi\bar
  \d c-4\bar\d \tilde\varphi\bar\d^2\tilde\varphi c
  -4(\bar\d\tilde\varphi)^2\bar \d c\\
  -\d^2\bar\d\bar c -2(\d\tilde\varphi \d+\d^2\tilde\varphi)\bar\d\bar
  c-2(\d\bar\d\tilde\varphi-\bar
  \d\tilde\varphi\d-2\bar\d\tilde\varphi \d\tilde\varphi ) \d\bar c
  \\-2(\d^2\bar\d\tilde\varphi+2\d\bar\d^2\tilde\varphi
  -2\bar\d\tilde\varphi\d^2\tilde\varphi
  -4\bar\d\tilde\varphi\d\bar\d\tilde\varphi )\bar c .
\end{gathered}
\end{equation}

Let us now set $\tilde\varphi=0$. Note that one can re-introduce an
arbitrary $\tilde\varphi$ through the finite coordinate transformation
generated by $\xi^u=-u\tilde\varphi$, $\xi^A=-\xi^u_{,B}\int ^\infty_r
dr^\prime (e^{2\beta} g^{AB})$, $\xi^r=-\half r(\d_A \xi^A
-2\tilde\varphi -f_{,B} U^B)$. The above relations then simplify to
\begin{equation}
\begin{split}
  \label{eq:30til}
  \beta&=-\frac{1}{4}r^{-2}c\bar c -\frac{1}{3}r^{-3} (d\bar
  c+\bar d c) +o(r^{-3-\epsilon}), \\
  U^\zeta&=-2r^{-2}\d \bar c-\frac{2}{3}r^{-3}\Big[(\ln
  r+\frac{1}{3})4\d\bar d-4 \bar c \bar \d c+ N^\zeta\Big]
  +o(r^{-3-\epsilon}),\\
  \frac{V}{r}&= r^{-1}2 M+o(r^{-1-\epsilon}),\\
  \d_u M&=\big[\d^2\dot{\bar c}+\bar \d^2\dot c\big] -\dot c\dot{\bar c},
  \\
  \d_u N_{\bar\zeta}&=
\bar \d M +\frac{1}{4}\big[(\bar\d \bar
  c+5\bar c\bar\d)\dot c-(3 c\bar\d+7\bar\d c) \dot{\bar c} \big]
+\bar \d^3c-\d^2\bar\d\bar c.
\end{split}
\end{equation}
When defining $\tilde M=M-\bar\d^2 c- \d^2 \bar c$ and $\tilde
N^\zeta=-\frac{1}{12}[2 N^\zeta+7\bar c \bar \d c+3 c\bar \d \bar c]$, the
evolution equations become
\begin{equation}
\begin{split}
  \label{eq:30q}
  \d_u\tilde M&=-\dot c\dot {\bar c},
  \\
  3\d_u \tilde N^\zeta&=-\bar\d \tilde M-2\bar \d^3  c-(\bar\d \bar c+3
  \bar c\bar \d )\dot{ c}.
\end{split}
\end{equation}

\subsection{Realization of $\mathfrak{bms}_4$ on solution space }
\label{sec:mass-asympt-flat}

In order to compute how $\mathfrak{bms}_4$ is realized on solution
space we need to compute the Lie derivative of the metric on-shell. We
will do so for the extended transformations defined by
\eqref{eq:4aext}-\eqref{eq:4bext} and use $-\delta
\bar\gamma_{AB}=2\omega\bar \gamma_{AB}$. Let
$\tilde\psi=\psi-2\omega$. This gives
\begin{equation}
  \label{eq:16}
  -\delta C_{AB}=[f\d_u+\cL_Y -\half(\tilde \psi+fl)] C_{AB}-2\bar D_A\bar D_B f+\bar \Delta f\bar\gamma_{AB},
\end{equation}
where \eqref{eq:99} should be used to eliminate $\d_u C_{AB}$ in favor
of $N_{AB}$ and
\begin{equation}
  \label{eq:19}
  -\delta D_{AB}=\cL_Y D_{AB}, 
\end{equation}
where we have used that 
\begin{equation*}
  \begin{split}
    \bar D_A\bar D_C f C^C_B+\bar D_B\bar D_C f C^C_A-
    \bar\gamma_{AB}\bar D_C\bar D_C f C^{CD}-\bar\Delta f C_{AB}=0,\\
    \bar D_A f \bar D_C C^C_B+\bar D_B f \bar D_C C^C_A+ \bar D_C f
    \bar D_A C^C_B+ \bar D_C f \bar D_B C^C_A-\\- \bar D^C f\bar D_C
    C_{AB}-\bar\gamma_{AB} \bar D_C f\bar D_D C^{CD} =0,
  \end{split}
\end{equation*}
which can be explicitly checked in the parametrization
$\bar\gamma_{AB}dx^Adx^B=e^{2\tilde\varphi} d\zeta d\bar \zeta$ with
$C_{AB}$ defined in \eqref{eq:17}.  By taking the time derivative of
\eqref{eq:16} and using \eqref{eq:99}, \eqref{eq:44ter} with $\psi$
replaced by $\tilde \psi$, one finds the transformation law for the
news tensor,
\begin{multline}
  \label{eq:22}
  -\delta N_{AB}= [f\d_u + \cL_Y] N_{AB}-(\bar D_A\bar D_B
  \tilde \psi-\half \bar \Delta\tilde \psi \bar\gamma_{AB})\\ 
+\frac{1}{4} (2 f\dot l
  +  f l^2+ \tilde \psi l -4\dot\omega+2
  Y^C\bar D_C l) C_{AB}\\+l (\bar D_A\bar D_B f-\half \bar \Delta
  f\bar\gamma_{AB}) -f(\bar D_A\bar D_B l -\half \bar\Delta l
  \bar\gamma_{AB}).
\end{multline}

We have $g_{uA}=\half \bar D_B C^B_A+\frac{2}{3} r^{-1} \big[(\ln
r+\frac{1}{3}) \bar D_B D^B_A+\frac{1}{4} C_{A}^B \bar D_C
C^{C}_B+N_A\big]+o(r^{-1-\epsilon})$, and by computing $\cL_\xi
g_{uA}$ on-shell, we find to leading oder that  
$-\delta (\bar D_B C^B_A)=[f\d_u+\cL_Y+\half(lf+\tilde\psi)] \bar D_B
C^B_A-\half \d_B (lf+\tilde\psi) C^B_A+\d_C
f(N^C_A+lC^C_A)-\d_A(\bar\Delta f)-\d_A f \bar R$. This is consistent
with \eqref{eq:16} by using the generalization of \eqref{eq:35} which reads
\begin{equation}
  \label{eq:35a}
  2\bar D_B\bar  D_C Y_A=\bar  \gamma_{CA}\bar  D_B\psi+\bar \gamma_{AB}
\bar D_C\psi-\bar\gamma_{BC}\bar  D_A \psi+\bar R
  Y_C \bar \gamma_{BA}-\bar R Y_A \bar \gamma_{BC},  
\end{equation}
and implies $\bar \Delta Y^A=-\half \bar RY^A$, $\bar\Delta \psi=-\bar
R\psi$. The logarithmic term gives 
$-\delta (\bar D_B D^B_A)=(f\d_u+\cL_Y+lf+\tilde\psi)\bar D_B D^B_A$,
which is again consistent with \eqref{eq:19}, while the $r^{-1}$
terms, when combined with the previous transformations, give
\begin{multline}
  \label{eq:33}
  -\delta N_A=[f\d_u +\cL_Y+\tilde\psi+fl]N_A -\half [ f\bar D_B l
  +\bar D_B \tilde \psi +(\tilde\psi+lf)
  \bar  D_B ] D^B_A\\
  +3\bar D_A f M -\frac{3}{16}\bar D_A f
  N^B_CC^C_B
+\half \bar D_B f N^B_C C^C_A
+\frac{1}{32}(  \bar D_A fl-f\bar D_A l-\bar D_A\tilde\psi)
  (C^B_CC^C_B)\\+ \frac{1}{4} (\bar D_Bf \bar R+\bar D_B \bar\Delta
  f)C^B_A -\frac{3}{4}\bar D_B f(\bar D^B\bar D_C C^C_A-\bar
  D_A\bar D_C C^{BC})
  \\+\half (\bar D_A\bar D_B f
-\frac{1}{2}\bar \Delta f\bar\gamma_{AB} )\bar D_C C^{CB}+\frac{3}{8} \bar
  D_A(\bar D_C\bar D_B f C^{CB}) . 
\end{multline}
Here $\d_u N_A$ should be eliminated by using 
\eqref{eq:78ter}. 
In the same way, from the order $r^{-1}$ of $\cL_\xi g_{uu}$, we get 
\begin{multline}
  \label{eq:34}
  -\delta M =[f\d_u+Y^A\d_A+\frac{3}{2} (\tilde\psi+
  fl)]M\\+\frac{1}{4}\d_u[
  \bar D_C \bar D_B f C^{CB} +2\bar D_B f\bar D_C C^{CB}]
+\frac{1}{4}[\bar D_A f l-f\bar D_A l-\bar D_A \tilde\psi]\bar D_B
C^{BA}
\\
+\frac{1}{4}\d_A
  f(\d^A\bar R-C^{AB}\bar D_B l)+\frac{1}{4}
  l[\bar D_C \bar D_B f C^{CB} +\bar D_B f\bar D_C C^{CB}], 
\end{multline}
where $\d_u M$ should be replaced by its expression from
\eqref{eq:3}. 

Let us now discuss these transformations in the parametrization
$\zeta,\bar\zeta$ with $\tilde\varphi=0=\omega$ so that
$\bar\gamma_{AB}dx^Adx^B=d\zeta d\bar\zeta$. From the leading and
subleading orders of $\cL_\xi g_{\zeta\zeta},\cL_\xi g_{\bar \zeta\bar
  \zeta}$, we get
\begin{equation}
  \label{eq:50}
  \begin{gathered}
-\delta c=f\dot c+Y^A\d_A c+(\frac{3}{2} \d  Y-\half  \bar \d 
\bar Y)c-2 \d^2 f  ,\\
-\delta d=Y^A\d_A  d+2\d Y d,
  \end{gathered}
\end{equation}
with $f$ given in \eqref{eq:48} and 
the complex conjugate relation holding for $\bar c, \bar d$. 
In particular, for the news function we find 
\begin{equation}
  \label{eq:28}
  -\delta\dot c=f\ddot c+Y^A\d_A \dot c+2 \d  Y\dot c-
  \d^3  Y  ,
\end{equation}
From the subleading term of
$\cL_\xi g^{r\zeta}$ and the leading term of $\cL_\xi g_{uu}$ and  we get
\begin{multline}
  \label{eq:59}
  -\delta \tilde N^\zeta=Y^A \d_A \tilde N^\zeta
  + (\d Y + 2\bar \d \bar Y )\tilde N^\zeta +\frac{1}{3} \d (\psi\bar d)\\
- \bar
  \d f (\tilde M +2
  \bar \d^2  c + \bar c \dot{ c} )-\frac{f}{3}\big[\bar\d \tilde M+2\bar \d^3
   c+(\bar\d  \bar c+3 \bar c\bar \d)\dot{ c}\big], 
\end{multline}
\begin{equation}
  \label{eq:58}
  -\delta \tilde M  = -f
   \dot c\dot{\bar c}+  Y^A \d_A \tilde M + \frac{3}{2}
  \psi \tilde M +\bar c \d^3 Y+  c \bar \d^3 \bar Y 
    +  4 \d^2 \bar \d^2 \tilde T. 
\end{equation}

As can be understood by comparing with the $3$ dimensional anti-de
Sitter and flat cases, this computation already contains information
on the central extensions in the surface charge algebra through the
inhomogeneous part of the transformation laws for the fields because
the normalization of the generators is known from the charges of the
Kerr solution. We plan to return to the details of this question
elsewhere.

\section{Conclusion and outlook}
\label{sec:conclusion-outlook}

In this work, we have shown that the symmetry algebra of
asymptotically flat $4$ dimensional spacetimes is $\mathfrak{bms}_4$,
an algebra that contains both the Poincar\'e algebra and the non
centrally extended Virasoro algebra in a completely natural way. As a
first non trivial effect, we have computed the transformation
properties of the data characterizing solution space. 

We have not analyzed in detail the singularities of the
$\mathfrak{bms}_4$ transformations nor those of the classical
solutions that ought to be allowed. Indeed, in the Bondi-Metzner-Sachs
gauge, the $\mathfrak{bms}_4$ transformations that we are advocationg
involve singularities at two points of the sphere at infinity that
extend to $\Im^+$ or to $\Im^-$ and into the bulk. The reason why we
have choosen this approach to asymptotically flat spacetimes at null
infinity is its similarity with the Fefferman-Grahan definition of
asymptotically anti-de Sitter spacetimes.  The appearance of the
extended transformations does not depend on this choice
however. Indeed, in the geometrical approach to asymptotic flatness
as developed in \cite{Geroch:1977aa}, there is also room for the
extended transformations since the quotient algebra of the asymptotic
symmetry algebra by the infinitesimal supertranslations is again
characterized by the conformal Killing vectors of the two-sphere.

More to the point, one can also consider a definition of asymptotic
flatness in 4 dimensions at null infinity that does not completely fix
the gauge, as done for instance in \cite{Brown:1986nw} for
asymptotically $AdS_3$ spacetimes. Such conditions can be inferred for
instance from equation (11.1.24) of \cite{Wald:1984rg}. In this case,
$\mathfrak{bms}_4$ appears as the quotient algebra
of allowed transformations modulo trivial ones. By adapting the
arguments of footnote 6 of \cite{Guica:2008mu} one can then improve
the asymptotic symmetry generators through pure gauge transformations
in such a way as to remove all singularities in the asymptotic
symmetry generators at finite radius. From this point of view, the
bulk singularities of the asymptotic symmetry generators appear as an
artefact of the Bondi-Metzner-Sachs gauge fixation.

We believe that our understanding of the symmetry structure and its
action on solution space goes some way in getting quantitative control
on ``structure X'' \cite{witten:98xx}, i.e., on a holographic
description of gravity with zero cosmological constant.

What we plan to do next is a systematic discussion of the central
extensions and the representation theory of $\mathfrak{bms}_4$ on the
one hand, and a construction of the associated algebra of surface
charges and generators on the other. In the future, it should be
interesting to analyze in more details the consequences of our results
on local conformal invariance for the non extremal Kerr/CFT
correspondence and for the gravitational S-matrix for instance.

\section*{Acknowledgements}
\label{sec:acknowledgements}

\addcontentsline{toc}{section}{Acknowledgments}

The authors thank M.~Ba\~nados, G.~Comp\`ere, G.~Giribet,
A.~Gomberoff, M.~Henneaux, A.~Kleinschmidt, C.~Mart\'{\i}nez,
R.~Troncoso and A.~Virmani for useful discussions. This work is
supported in parts by the Fund for Scientific Research-FNRS (Belgium),
by the Belgian Federal Science Policy Office through the
Interuniversity Attraction Pole P6/11, by IISN-Belgium and by Fondecyt
projects No.~1085322 and No.~1090753.




\def\cprime{$'$}
\providecommand{\href}[2]{#2}\begingroup\raggedright\endgroup

\end{document}